\documentclass[a4paper,fleqn]{cas-dc}
\usepackage[numbers]{natbib}
\usepackage{amsfonts}
\usepackage{algorithmic}
\usepackage{amsmath} 
\usepackage{amssymb} 
\usepackage{caption}
\usepackage{subcaption}
\usepackage{etoolbox}
\usepackage{forest}
\usepackage{lingmacros}
\usepackage{textcomp}
\usepackage{tree-dvips}
\usepackage{tikz}
\usepackage{tikz-cd}
\usepackage[arrowdel]{physics}
\usepackage{graphicx}
\usepackage{wrapfig}
\usepackage{listings}
\usepackage{pgfplots, pgfplotstable}
\usepackage{diagbox} 
\usepackage[usestackEOL]{stackengine}
\usepackage{makecell}
\usepackage{mathrsfs}
\usepackage{moresize}
\usepackage{multirow}
\usepackage{multicol}
\usepackage[numbers]{natbib}
\usepackage[T1]{fontenc}
\usepackage{xcolor}
\allowdisplaybreaks[1]
\definecolor{orchid}{rgb}{0.7, 0.4, 1.1}
\definecolor{comment_color}{rgb}{0, 0.5, 0}
\definecolor{keyword_color}{rgb}{0.3, 0, 0.6}
\definecolor{string_color}{rgb}{0.5, 0, 0.1}

\begin{document}
\shorttitle{Electromagnetic analysis of low dropout regulator circuit with small-signal stability characterization for inductive LC filters and transformers}

\shortauthors{Xi Liu$^a$, Wenxi Fang$^b$, Ken Perlin$^c$}

\title{{\Large Electromagnetic analysis of low dropout regulator circuit with small-signal stability characterization for inductive LC filters and transformers}}

\author[1]{\color{black}Xi Liu}
\author[2]{\color{black}Wenxi Fang}
\author[3]{\color{black}Ken Perlin}

\address{$^a$xl3467@columbia.edu, Columbia University; $^b$u3013972@connect.hku.hk, Inner Mongolia University of Science and Technology;
$^c$perlin@nyu.edu, New York Unversity}


\begin{abstract}
This work systematically evaluates the capability of generative large language models (LLMs), specifically GPT-4o, to support the full design, simulation, and optimization workflow of low-dropout (LDO) linear voltage regulators. The study covers four core design phases: pre-design specification mapping, transistor-level circuit topology generation, SPICE simulation guidance, and post-simulation performance fine-tuning, with an extended investigation into the integration of magnetic inductive components within the LDO signal path. GPT-4o autonomously proposes a single-stage differential-pair error amplifier architecture with thin-oxide MOS transistors, provides sizing guidance for the PMOS pass element, and recommends passive compensation networks to secure closed-loop stability. The LDO testbench adheres to low-voltage portable electronics specifications: an input range of 0.8-1.2 V, tunable 0.7-1.1 V output, maximum 250 mA load current, and integrable output capacitance below 10 nF. Transient and small-signal AC SPICE simulations validate LLM-assisted circuit implementations, quantifying settling time reduction via compensation capacitors and verifying adequate phase margin across operating bandwidth. A key novel extension explores three distinct inductor placement schemes, input-side supply filtering, out-of-loop LC output filtering, and inductive loading embedded within the feedback divider, rooted in Maxwell's electrodynamic principles and MOSFET small-signal device physics. Comparative Bode and output impedance analysis reveals that inductors inserted inside the feedback path introduce resonant complex poles, severe gain peaking, and degraded phase margin, while inductors placed external to the feedback sensing tap preserve regulator stability while suppressing high-frequency electromagnetic interference. Despite robust performance in topology drafting and simulation instruction, GPT-4o exhibits notable limitations: it occasionally omits critical passive components matching design constraints, generates syntactically flawed SPICE netlists for custom transistor subcircuits, and fails self-correction for unconventional magnetic load configurations. Overall, this work demonstrates that LLMs function as powerful auxiliary engineering assistants to accelerate iterative analog design, though rigorous human cross-verification of component sizing, loop stability, and passive network topology remains mandatory for reliable LDO implementation, particularly when integrating inductive magnetic elements.
\end{abstract}

\begin{keywords}
large language model\sep
low dropout regulator\sep
power management
\end{keywords}
\maketitle
\section{Introduction}
Complementing the core analog design challenges of LDO regulators, the integration of magnetic inductive passive elements introduces critical electromagnetic tradeoffs that demand rigorous first-principles analysis, a topic minimally addressed in existing LLM-assisted power circuit literature. From fundamental Maxwell equations, Faraday’s law of induction defines the frequency-dependent impedance of lumped inductors, which act as near-short circuits for DC steady-state operation yet inject significant phase lag at high AC frequencies when embedded within closed feedback signal paths. For low-voltage integrable LDOs constrained to small sub-10 nF output capacitors, series LC filter networks formed by surface-mount inductors and miniature output capacitors create second-order resonant pole pairs whose location relative to the feedback sensing tap dictates overall loop stability. Inductors placed at the input supply or directly on the load side, downstream of the feedback voltage divider, only modify supply and load impedance characteristics without altering the regulator’s core open-loop transfer function. Conversely, inductive loading inserted into the feedback resistor divider couples magnetic flux dynamics directly into the error amplifier’s sensing signal, degrading phase margin, triggering gain peaking, and generating severe transient voltage ringing. While prior LDO research has extensively covered capacitive frequency compensation and op-amp small-signal transistor physics, few generative AI design workflows systematically compare the stability impacts of multiple discrete inductor placement topologies. This work therefore extends standard LLM-guided LDO design practice by incorporating quasi-static magnetostatic modelling, LC damping factor optimisation, and comparative AC small-signal characterisation of three distinct magnetic circuit configurations, bridging electromagnetic physics, linear regulator stability theory, and transformer-based circuit automation.

The growth of transformers has empowered people with capabilities that surpass what we once thought possible, particularly through large language models (LLMs). OpenAI's GPT-4o are now capable of doing things that require expertise. The hope is that if these models can help perform complex tasks, they have the potential to enable time savings and better workflows, and even lower the barrier to expert insights. Research has shown that AI techniques, such as reinforcement learning and graph neural networks, can optimize analog and mixed-signal circuit designs \cite{turky_1989}, including LDO regulators, by automating parameter tuning and topology selection. For example, Li and Carusone \cite{lizonghao_2023} demonstrated the use of relational graph neural networks to optimize LDO designs in the open-source Sky130 process, achieving enhanced performance metrics through automated iteration. Similarly, advancements in deep learning have enabled automated circuit synthesis \cite{settaluri_2020}, as illustrated by approaches leveraging generative networks to produce novel designs in response to high-level specifications.

This project explores how useful a ChatGPT-4o can be during the creation of an Analog LDO design. Analog LDOs are used to maintain a noise-free voltage supply and provide the required stable supply for analog circuits. Typically, the LDO design process is an iterative endeavor that requires experience and extensive experimentation. With the introduction of ChatGPT-4o we are hoping to see if it can speed up this process, provide some sort of insight, and even highlight early errors.

The objective is to understand the different stages of the Analog LDO design process where ChatGPT-4o can be useful. More specifically, we would like to know its potential in the following areas: navigating us through design choices; how to narrow down options; what algorithms could fit as well as possible design pitfalls. We would like to know if LLMs such as ChatGPT-4o could turn into useful engineering assistants, not necessarily just towards generating any individual design but in enhancing its process overall when asked these questions.

Here, we shall describe how we employed ChatGPT-4o in the Analog LDO design process and utilized data from it to shed light on the parameters and our results with it through this paper. We will analyze the successes and failures of the model, before discussing what this could reveal for future AI in circuit design.

The rising demand for portable electronics has heightened the need for low-voltage, low-dropout (LDO) regulators. These components are frequently paired with DC-DC converters or used independently, often in conjunction with switching regulators to minimize noise and deliver a stable output \cite{rincon_1998}. As portable devices evolve toward lower operating voltages due to shrinking feature sizes and reduced breakdown voltages, minimizing quiescent current becomes increasingly critical for extending battery life in these systems.

\section{Conversation-based model construction}
\subsection{Design specifications}
The design specifications for the low dropout regulator are:
Input voltage range: 0.8 $\sim$ 1.2 V.
Output voltage: 0.7 $\sim$ 1.1 V (dropout voltage = 0.1 V).
For example: In the case Vin=1.2V, Vout should cover 0.8 $\sim$ 1.1 V.
Maximum load current: 250 mA.
Output capacitor: $<$ 10 nF (Integrable).
Thin-oxide devices preferred.

\subsection{Circuit design using simulation program with integrated circuit emphasis (SPICE)}
Analog simulations typically use simulation programs with integrated circuit emphasis (SPICE) models or similar tools that can model transistor-level behavior and physical device properties. Designing an LDO often requires mixed-signal modeling, combining analog components like the pass transistor with some digital control logic \cite{lizonghao_2023}. Tools such as SPICE simulators (e.g., Cadence Spectre, LTspice) are appropriate for LDO design.

\begin{figure}
\centering
\includegraphics[width = 0.4\textwidth, height = 0.15\textwidth]{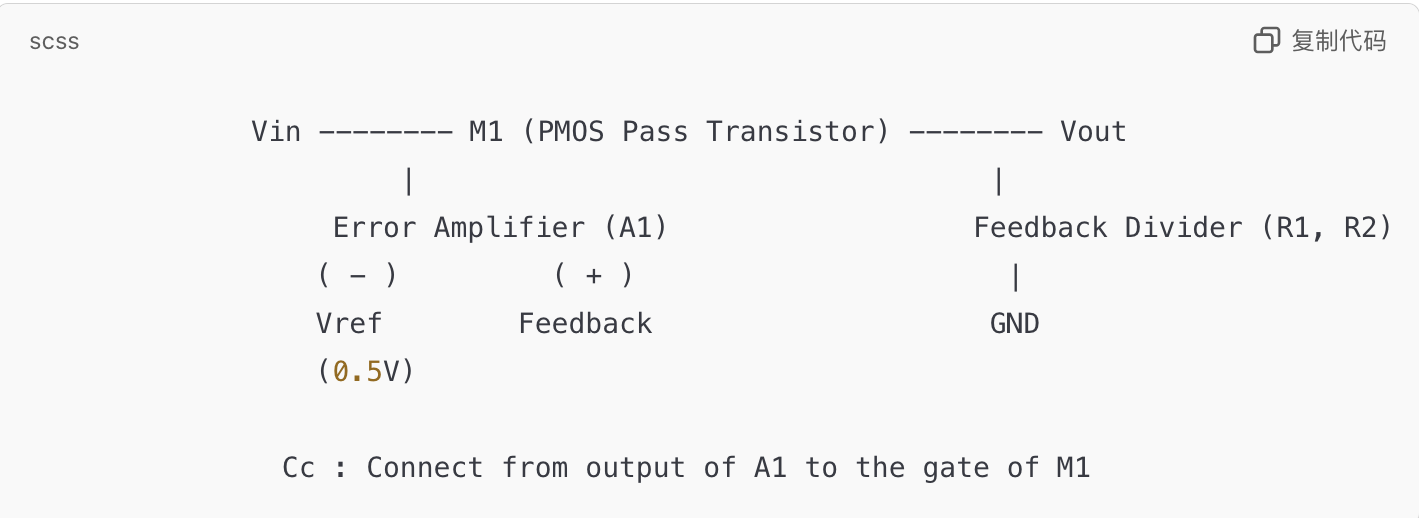}
\caption{Model schematic provided by ChatGPT-4o}
\label{model construction ChatGPT schematic}
\end{figure}

To construct the model of the LDO design based on the conversation with ChatGPT, we first need to provide the design specifications and ask for the specific schematic design. Figure \ref{model construction ChatGPT schematic} shows how the ChatGPT generates the LDO schematic, which includes the components that are needed to build the LDO circuit diagram and the position of the connection of the specific component, for example, the compensation capacitor (Cc).

\begin{figure}
\centering
\includegraphics[width = 0.45\textwidth, height = 0.3\textwidth]{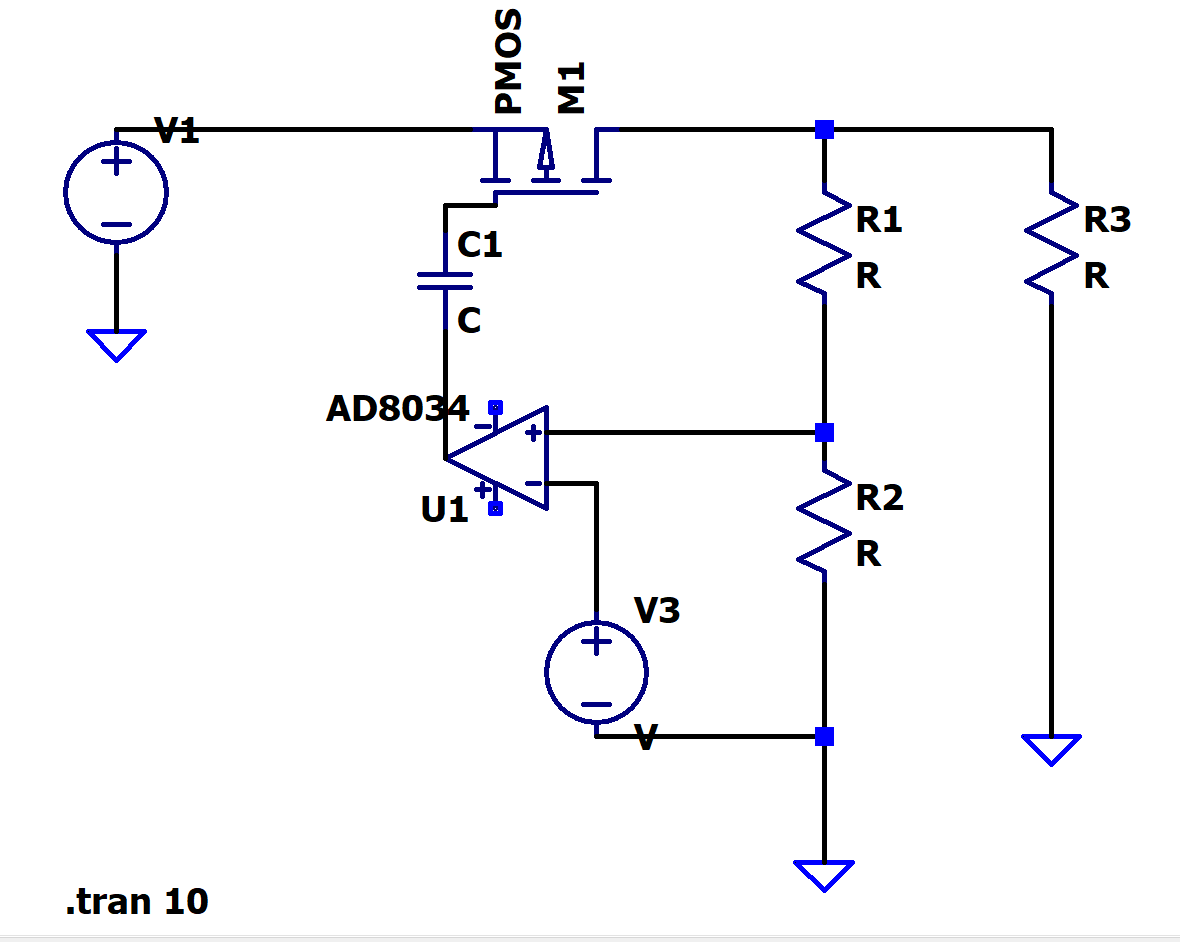}
\caption{Model construction on LTspice before correctness check}
\label{model construction ltspice}
\end{figure}

Although ChatGPT can not directly generate the LTspice circuit diagram, we can still ask ChatGPT to provide the detailed steps to help us re-create the LDO circuit in LTspice. Then we followed the design, built the circuit in LTspice shown in Figure \ref{model construction ltspice}, and sent it back to ChatGPT to check the correctness. 

\begin{figure}
\centering
\includegraphics[width = 0.45\textwidth, height = 0.3\textwidth]{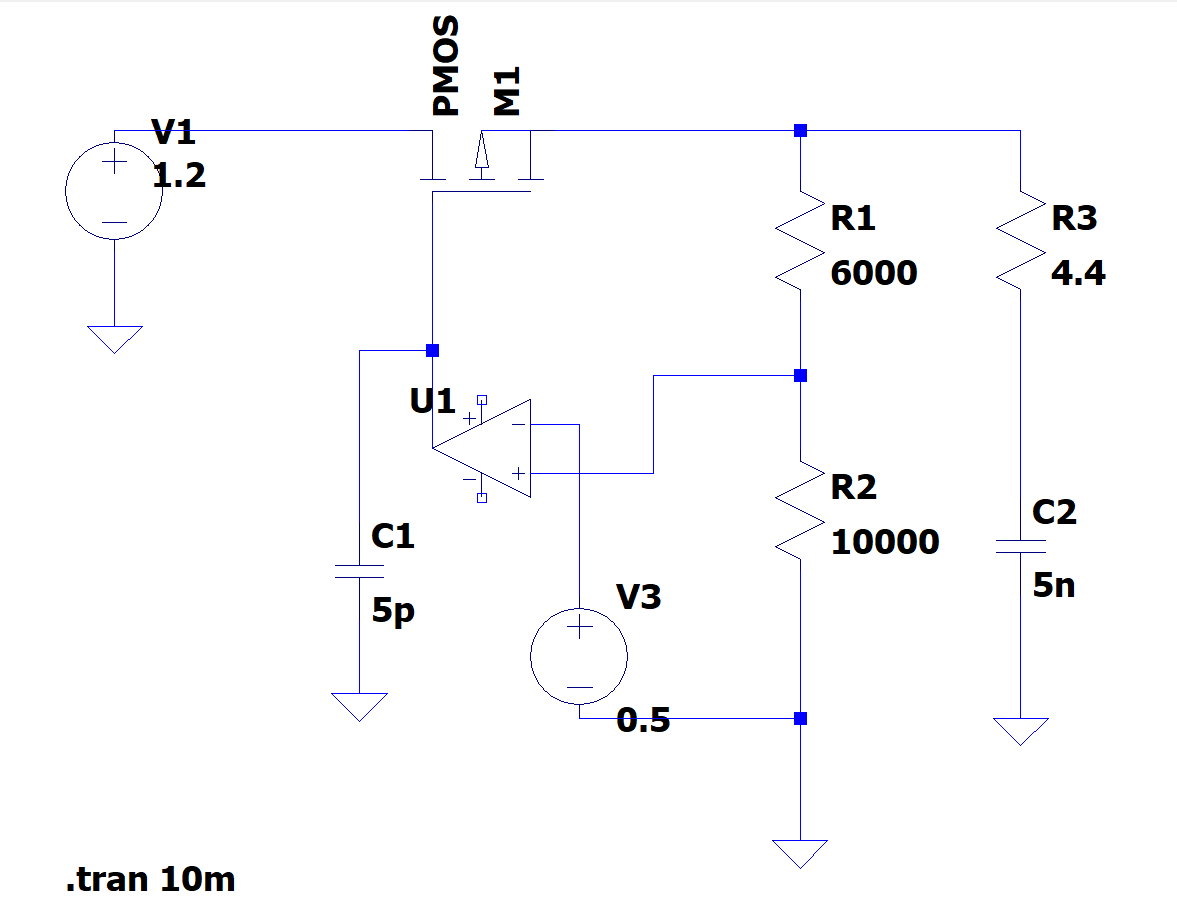}
\caption{Final Model construction on LTspice}
\label{Final model construction ltspice}
\end{figure}

The ChatGPT correctness check confirmed that our circuit design is in accordance with the schematic provided by ChatGPT. However, according to the self-check, ChatGPT did not realize and indicate what the LDO circuit design was missing. In the model schematic provided by ChatGPT-4o, ChatGPT loses sight of the initial requirement for the component output capacitor, which is less than 10 nF. Therefore, we feed the self-check result back to ChatGPT to ask for the new schematic of the circuit connection and re-create the LDO circuit shown in Figure \ref{Final model construction ltspice}.

After the correctness check of the final model construction shown in Figure \ref{Final model construction ltspice}, we ask ChatGPT to assign specific data to each component based on the design specifications. To satisfy the required output voltage range, ChatGPT assigns the input voltage V1 = 1.2 V, the reference voltage source V3 = 0.5 V, feedback resistors R1 and R2 = 6k Omega and 10k Omega, and load resistor R3 = 4.4 Omega. 

ChatGPT did not mention the specific model of the error amplifier. The error amplifier (EA) we chose is an AD8034 amplifier with thin-oxide devices, preferred for speed and integrability. We then asked GPT to validate the decision and got positive feedback. The input of the amplifier is tied to the feedback voltage from the output through a resistor divider, and the other input is connected to a reference voltage \( V_{\text{ref}} \), which is generated by a bandgap or other reference source. The output of the error amplifier drives the gate of the pass element, which is a PMOS transistor \cite{leung_2003}. The PMOS transistor serves as the series pass element, with its source connected to \( V_{\text{in}} \) and its drain connected to \( V_{\text{out}} \). The gate is controlled by the output of the error amplifier.

In the circuit, a resistor divider made of resistors \( R_1 \) and \( R_2 \) is connected between \( V_{\text{out}} \) and ground, providing the feedback voltage \( V_{\text{fb}} = V_{\text{out}} \times \frac{R_2}{R_1 + R_2} \). This feedback voltage is then compared to the reference voltage \( V_{\text{ref}} \) by the error amplifier. The maximum load current is determined by the sizing of the PMOS pass element, it is checked whether it can handle the required 250 mA load \cite{anusha2020}. The compensation network for stability includes a small compensation capacitor \( C_c \) (figure \ref{LDO with capacitor $C_2$ PULSE(0.8 1.2 0 1u 1u 5m 10m)}), typically in the range of a few picofarads, placed between the error amplifier output and ground. The output capacitor \( C_{\text{out}} \), which should be less than 10 nF, is used to ensure stability and proper regulation. The reference voltage \( V_{\text{ref}} \), often a stable bandgap or trimmed reference voltage (e.g., 0.5 V), serves as the reference for the error amplifier \cite{kao2024feedforward}. The expected circuit schematic would resemble the LDO topology in the uploaded image, but with adjustments such as a resistor divider for feedback, a reference voltage source for the error amplifier, and the addition of the compensation capacitor \( C_c \) for stability.


\begin{figure}
\centering
\begin{subfigure}{0.23\textwidth}
    \centering
    \includegraphics[width=\textwidth, height=0.8\textwidth]{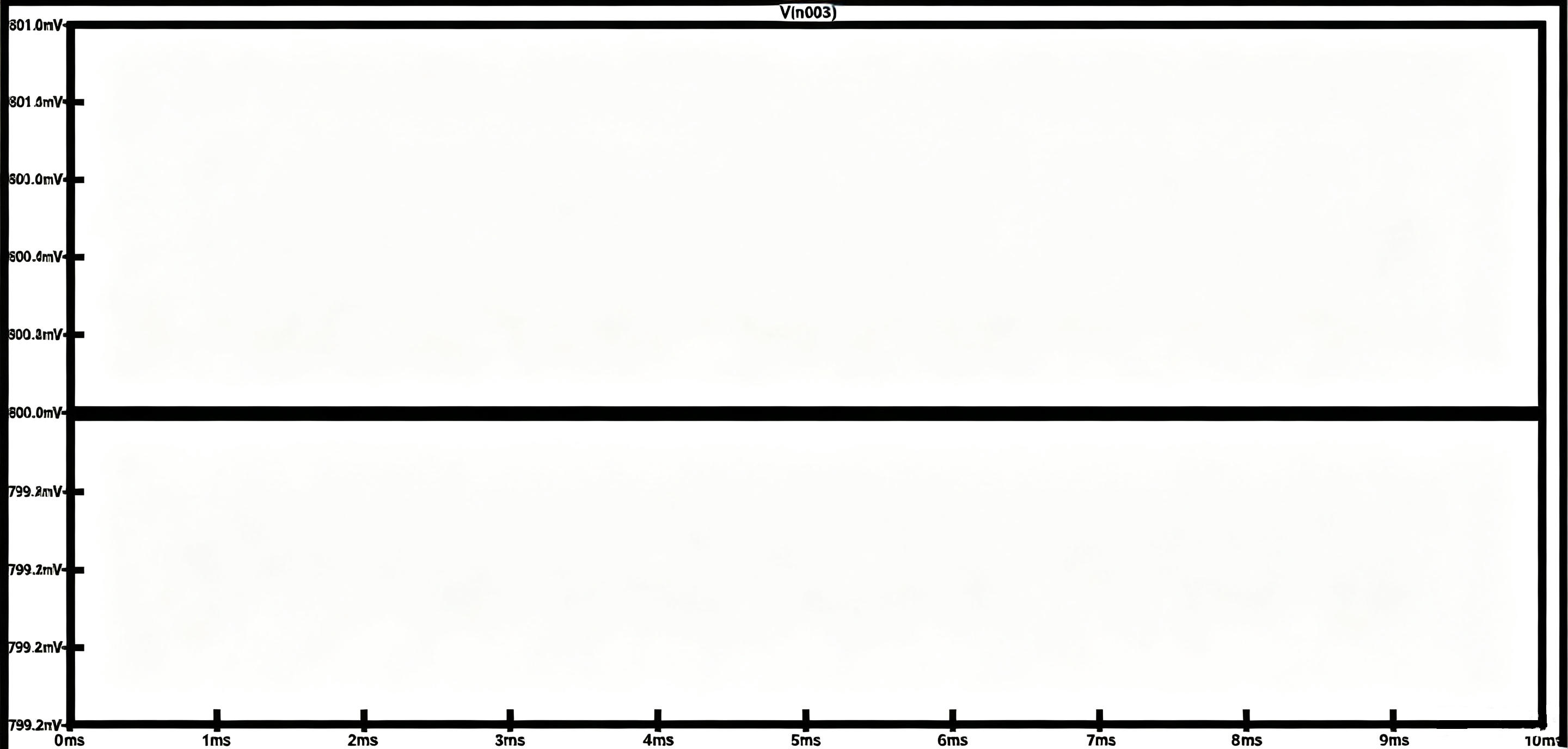}
    \caption{Output voltage given by 1.2V input}
    \label{simple ldo, voltage at output}
\end{subfigure}
\begin{subfigure}{0.23\textwidth}
    \centering
    \includegraphics[width=\textwidth, height=0.8\textwidth]{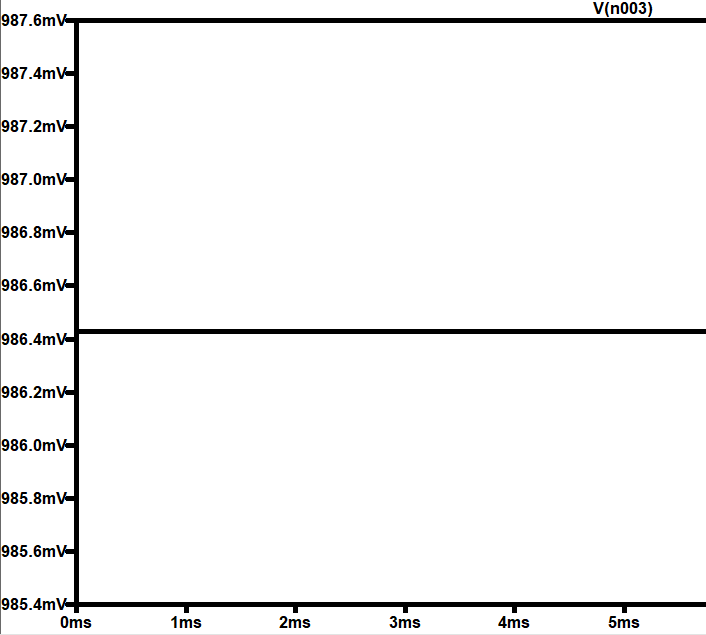}
    \caption{Output voltage, LDO with single stage amplifier}
    \label{output voltage, single stage amplifier}
\end{subfigure}
\caption{Output voltage comparisons}
\label{fig:vout_comparison}
\end{figure}

An input voltage of 1.2V is used for the designed circuit, the steady output is 1.11V as shown in Figure \ref{simple ldo, voltage at output}, the output voltage is 0.8V, which meets the design specifications.

\subsection{PULSE Simulation}
\begin{figure}
\centering
\includegraphics[width = 0.45\textwidth, height = 0.3\textwidth]{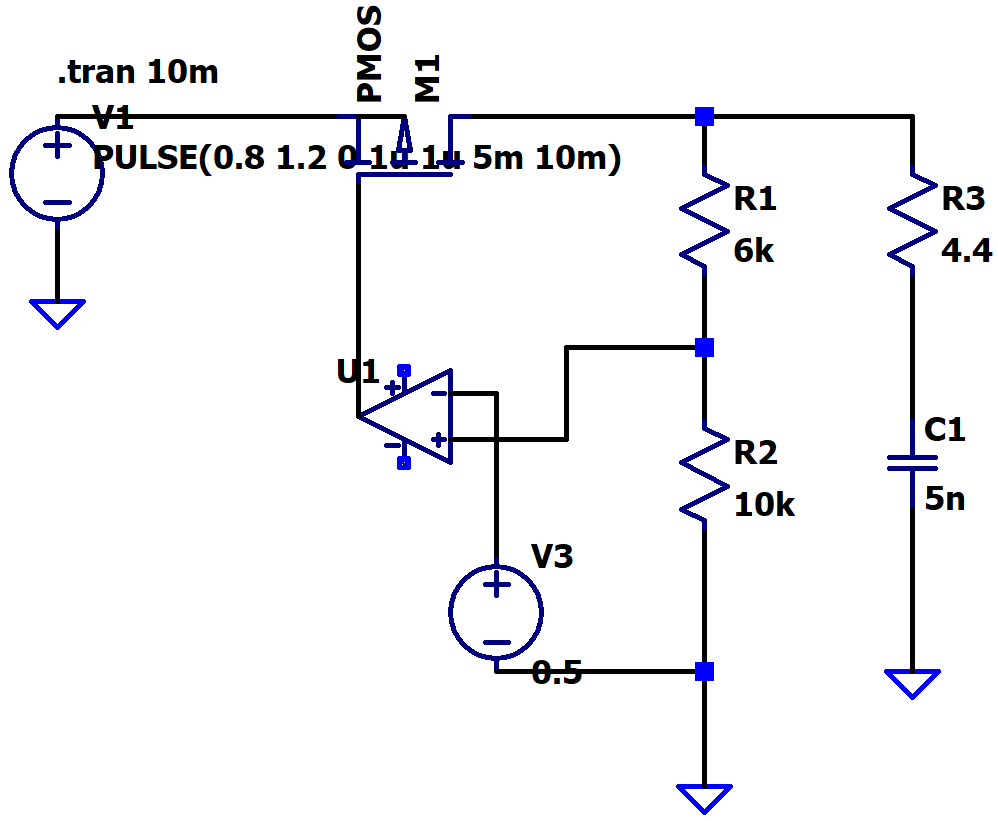}
\caption{LDO with PULSE(0.8 1.2 0 1u 1u 5m 10m)}
\label{LDO with PULSE(0.8 1.2 0 1u 1u 5m 10m)}
\end{figure}

\begin{figure}
\centering
\includegraphics[width = 0.4\textwidth, height = 0.27\textwidth]{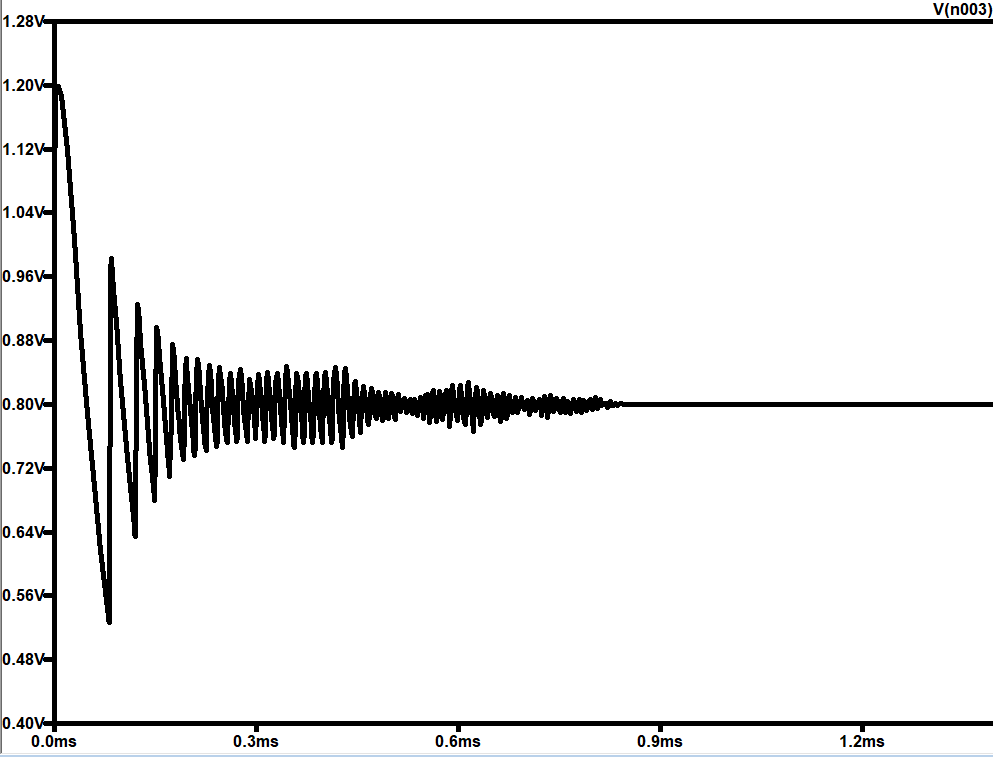}
\caption{voltage vs time graph for LDO with PULSE(0.8 1.2 0 1u 1u 5m 10m)}
\label{voltage vs time graph for LDO with PULSE(0.8 1.2 0 1u 1u 5m 10m)}
\end{figure}

The PULSE function in the LDO circuit is used to define a time-varying input voltage source \cite{zhan2024}. See figure \ref{LDO with PULSE(0.8 1.2 0 1u 1u 5m 10m)} for the LDO circuit with the pulse function settings, figure \ref{voltage vs time graph for LDO with PULSE(0.8 1.2 0 1u 1u 5m 10m)} is the voltage vs time graph for LDO with PULSE(0.8 1.2 0 1u 1u 5m 10m). It generates a periodic pulse waveform that alternates between a minimum voltage (\( V_1 \)) and a maximum voltage (\( V_2 \)) with specific timing parameters.

\begin{align*}
&PULSE(V1=0.8, V2=1.2, Tdelay=0, Trise=1u,\\
&Tfall=1u, Ton=5m, Tperiod=10m, Ncycles)
\end{align*}
In this case, \( V_1 = 0.8 \, \text{V} \) represents the lower end of the input voltage range, while \( V_2 = 1.2 \, \text{V} \) represents the upper end. These voltages are chosen to test the LDO's performance across its specified input range. The parameters \( T_{\text{rise}} = 1 \, \mu\text{s} \) and \( T_{\text{fall}} = 1 \, \mu\text{s} \) define the smooth voltage transitions during the rising and falling edges, ensuring the simulation accurately reflects realistic conditions.

The pulse remains at \( V_2 = 1.2 \, \text{V} \) for \( T_{\text{on}} = 5 \, \text{ms} \), testing the LDO's ability to regulate the output voltage under a high input voltage for an extended time. The total period of one cycle is \( T_{\text{period}} = 10 \, \text{ms} \), meaning the pulse alternates between \( V_1 \) and \( V_2 \) every 10 milliseconds. This periodic behavior evaluates the dynamic response of the LDO to input voltage variations and its ability to maintain stability and regulate the output voltage effectively.

In the LDO circuit, the PULSE function is critical for validating the regulator's transient and steady-state performance. We will discuss the GPT's capability of instructing on analysis guidance. When the input voltage rises to \( V_2 = 1.2 \, \text{V} \), the LDO should maintain the output within the specified range (\( 0.7 \sim 1.1 \, \text{V} \)), while ensuring minimal dropout and stability. Conversely, when the input voltage drops to \( V_1 = 0.8 \, \text{V} \), the LDO must still regulate the output properly, even at the lower limit of operation. The transitions test the transient response, ensuring there are no significant overshoots or undershoots, while the on-time and period evaluate the LDO’s performance under steady-state conditions.


\begin{figure}
\centering
\includegraphics[width = 0.45\textwidth, height = 0.3\textwidth]{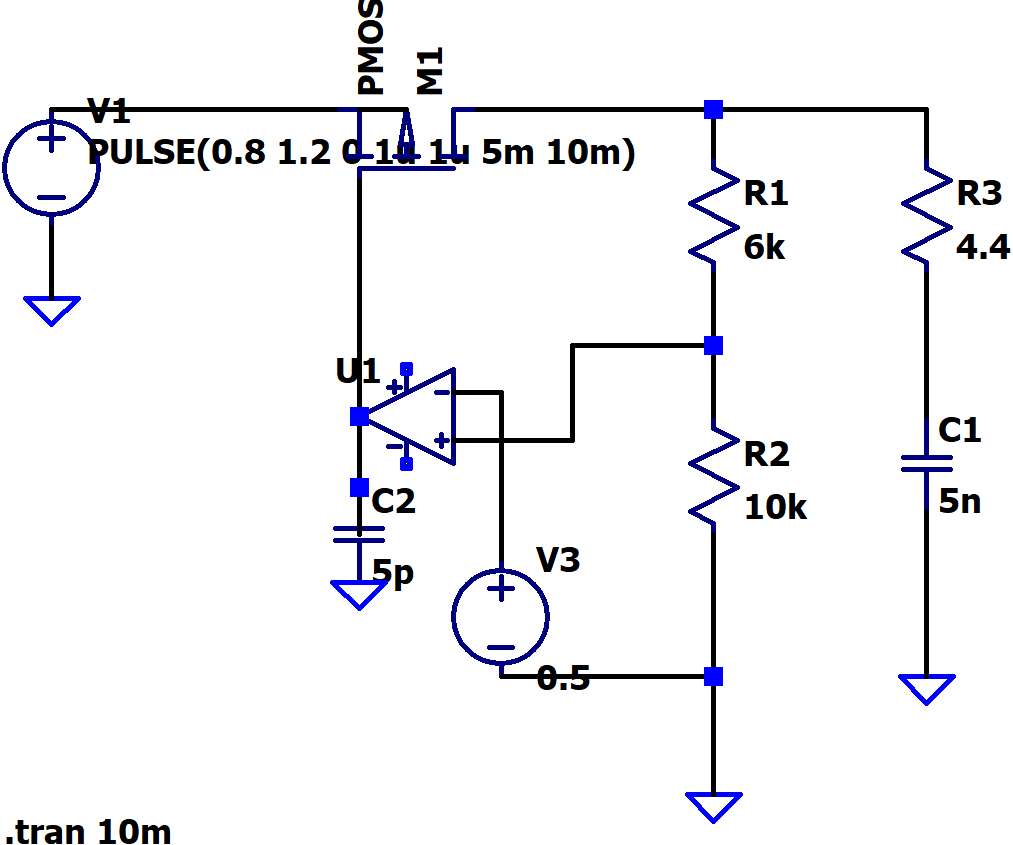}
\caption{LDO with capacitor $C_2$ with PULSE(0.8 1.2 0 1u 1u 5m 10m)}
\label{LDO with capacitor $C_2$ PULSE(0.8 1.2 0 1u 1u 5m 10m)}
\end{figure}

ChatGPT suggested the addition of a capacitor $C_2$ for stability. We then measured the effect of adding this capacitor. The capacitor \( C_2 \) affects the stability and response time of the circuit \cite{roshna_2024}. When \( C_2 \) is added to the circuit, it provides a bypass or compensation path that smooths out rapid fluctuations in voltage during the transient phase. \( C_2 \) works in conjunction with the resistors and active elements (e.g., transistors) to modify the frequency response and damping of the system (figure \ref{LDO with capacitor $C_2$ PULSE(0.8 1.2 0 1u 1u 5m 10m)}). This improves how the circuit transitions to steady state after changes in input conditions. Without \( C_2 \), the circuit may exhibit a slower response to input voltage changes due to the absence of this compensation \cite{kejia_2026}. The absence of \( C_2 \) leaves the circuit more susceptible to oscillations or ringing, as it relies only on the intrinsic properties of the error amplifier and pass element to stabilize. So it takes longer to reach the steady state, as the system requires additional time to settle around \( 0.8 \, \text{ms} \).

\begin{figure}
\centering
\includegraphics[width = 0.4\textwidth, height = 0.26\textwidth]{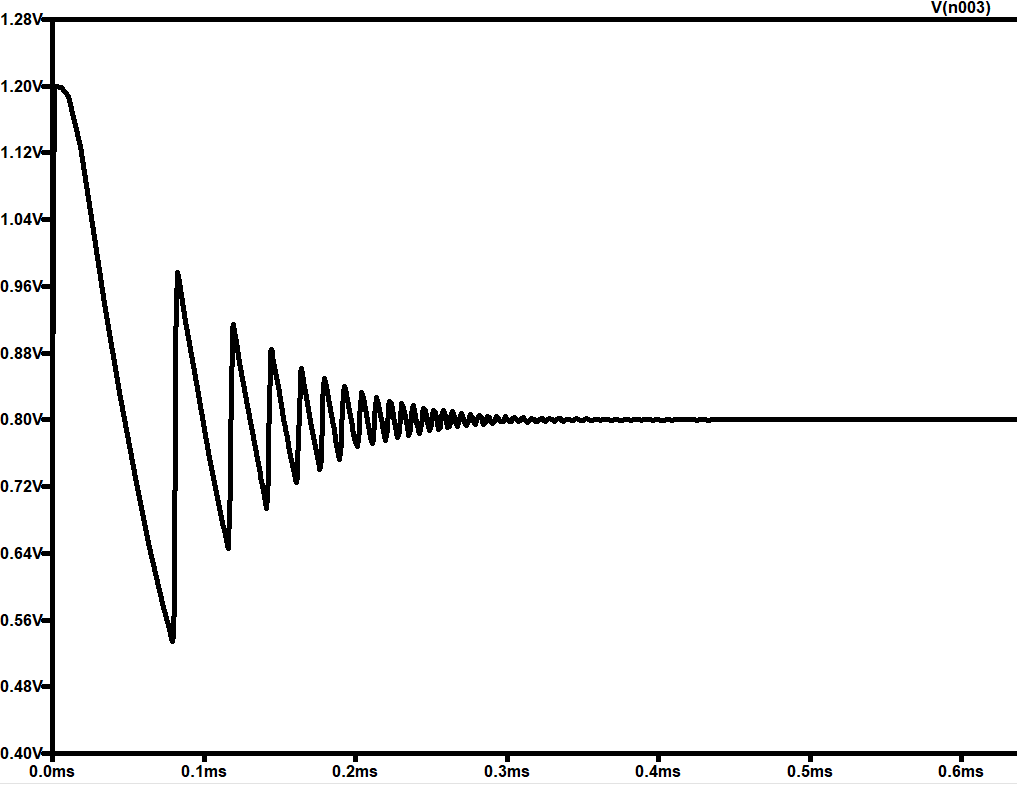}
\caption{voltage vs time graph for LDO with capacitor $C_2$ with PULSE(0.8 1.2 0 1u 1u 5m 10m)}
\label{voltage vs time graph for LDO with capacitor $C_2$ PULSE(0.8 1.2 0 1u 1u 5m 10m)}
\end{figure}

With \( C_2 \) added, the capacitor introduces a pole or zero into the system's transfer function, which can dampen oscillations and increase the phase margin. This enhanced damping reduces overshoot and allows the circuit to converge to its final output voltage more quickly (figure \ref{voltage vs time graph for LDO with capacitor $C_2$ PULSE(0.8 1.2 0 1u 1u 5m 10m)}). The improved response is evident in the reduction of the steady-state time to \( 0.3 \, \text{ms} \). \( C_2 \) filters high-frequency noise and transient disturbances, further contributing to faster stabilization. The capacitor \( C_2 \) improves the dynamic performance of the circuit by smoothing out transients, reducing oscillations, and increasing damping, which together allow the system to reach steady state more quickly.

\begin{figure}
\centering
\includegraphics[width = 0.4\textwidth, height = 0.26\textwidth]{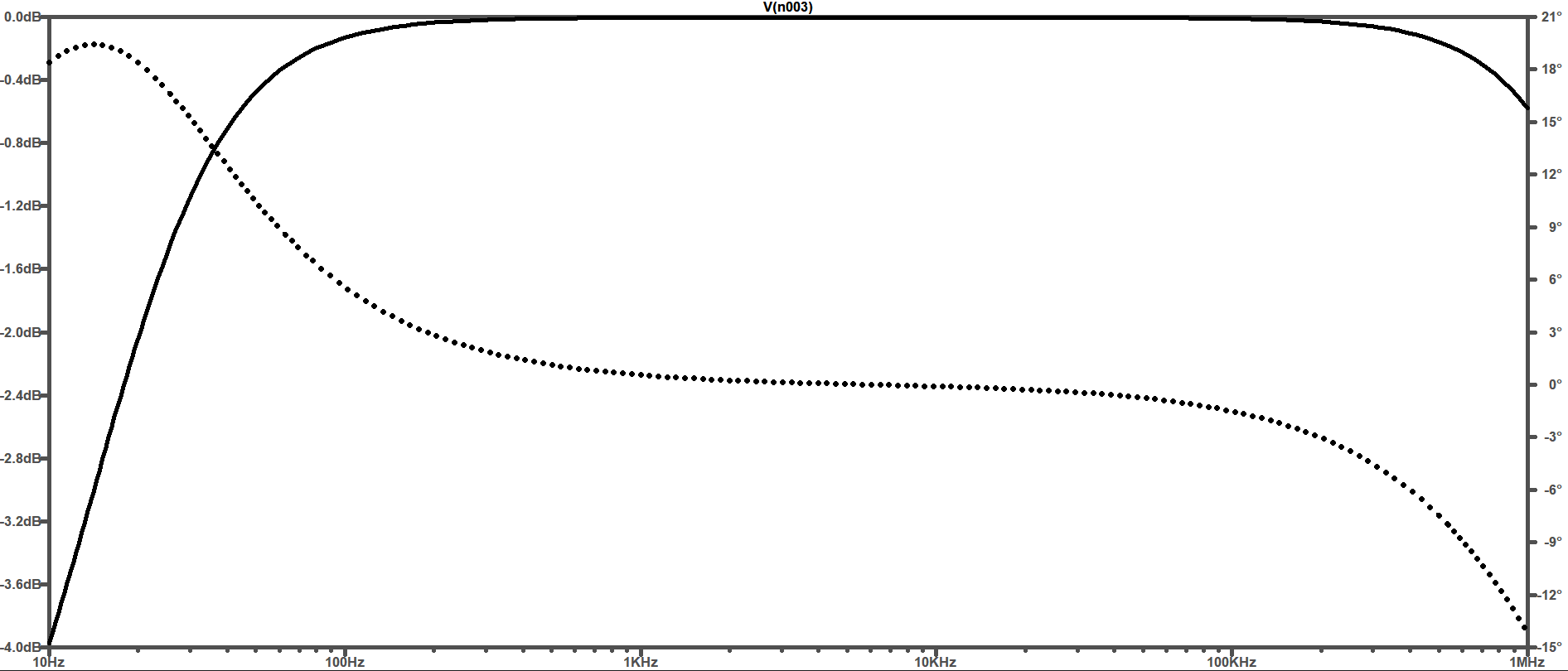}
\caption{LDO frequency response graph}
\label{LDO frequency response graph}
\end{figure}

The .ac analysis command in SPICE performs a small-signal AC analysis. The `.ac dec 100 10 1meg' command specifies a frequency sweep in decades, starting from \( 10 \, \text{Hz} \) to \( 1 \, \text{MHz} \), with \( 100 \) points per decade. The AC analysis evaluates the circuit's response to small AC signals at a range of frequencies, providing detailed data over several orders of magnitude.

The small-signal AC amplitude of \( 1 \) indicates that a unit amplitude signal is applied to the circuit, and its response is measured (figure \ref{LDO frequency response graph}). The curves on the \( \text{dB} \) vs. frequency graph correspond to the gain response and impedance response. The decreasing curve represents the gain response of the open-loop transfer function of the LDO circuit. The gain starts at a high value at low frequencies, representing the DC gain of the error amplifier and the pass element. As frequency increases, the gain decreases due to the introduction of poles in the system. The increasing curve represents the impedance or load regulation response of the LDO circuit. At low frequencies, the impedance is high, and as frequency increases, the impedance decreases, eventually approaching an asymptotic value. This behavior reflects the frequency-dependent feedback loop and the effect of bypass capacitors like \( C_2 \).

The analysis evaluates the stability, bandwidth, and load regulation of the LDO circuit. The gain response determines stability metrics such as the phase margin, gain margin, and unity-gain bandwidth. Stability is ensured if the phase margin at the unity-gain frequency is adequate, typically greater than \( 45^\circ \). The impedance response evaluates the effectiveness of the LDO's output voltage regulation under dynamic load conditions, ensuring that the output impedance remains low across the frequency range and minimizing the impact of load transients. This analysis ensures proper operation of the LDO under various frequency-dependent conditions.

\section{Conversation-based Analysis guidance}
\begin{figure}
\centering
\begin{subfigure}{0.23\textwidth}
\centering
\includegraphics[width=\textwidth, height=0.8\textwidth]{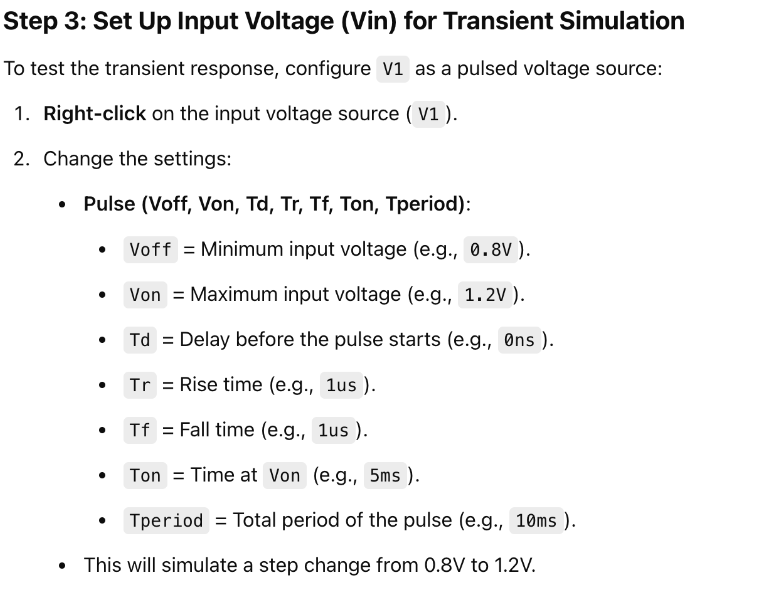}
\caption{Vin Configuration}
\label{Transient Analysis Vin Configuration by GPT-4o}
\end{subfigure}
\begin{subfigure}{0.23\textwidth}
\centering
\includegraphics[width=\textwidth, height=0.8\textwidth]{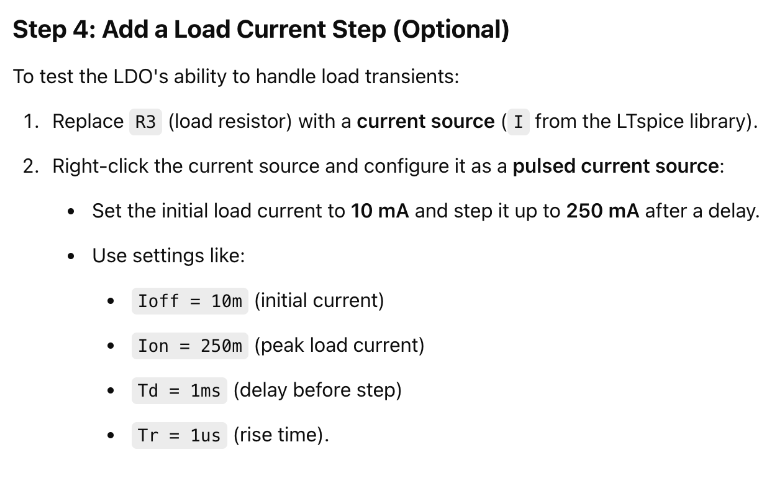}
\caption{Optional Step}
\label{Optional Analysis Step by GPT-4o}
\end{subfigure}
\begin{subfigure}{0.23\textwidth}
\centering
\includegraphics[width=\textwidth, height=0.6\textwidth]{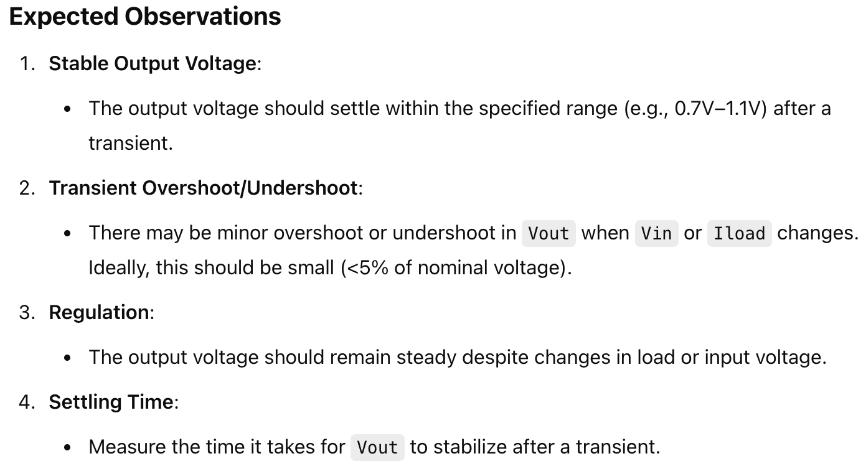}
\caption{Expected Observations}
\label{Expected Observations by GPT-4o}
\end{subfigure}
\begin{subfigure}{0.23\textwidth}
\centering
\includegraphics[width=\textwidth, height=0.6\textwidth]{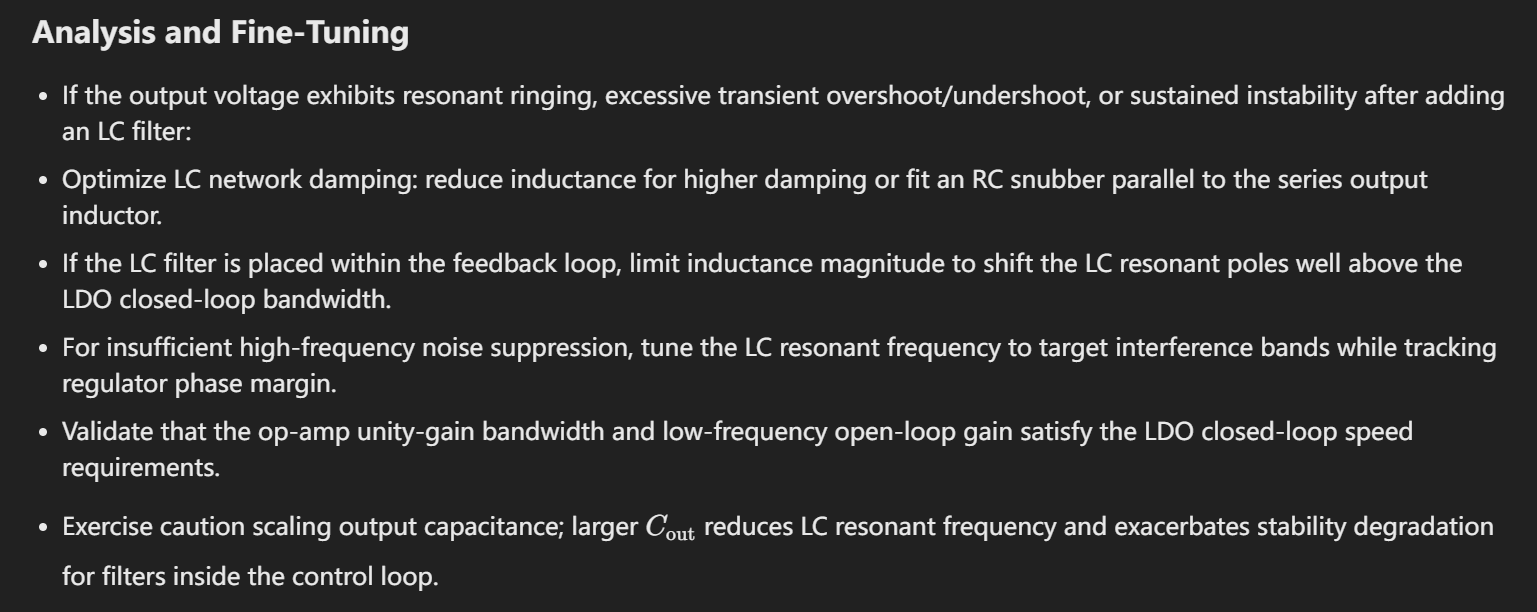}
\caption{Fine Tuning Suggestions}
\label{fig:fine_tuning}
\end{subfigure}
\caption{Transient Analysis Configuration and Observations from GPT-4o}
\label{Fine Tuning Suggestions by GPT-4o}
\end{figure}

To explore the capability of GPT-4o on instructing analyses. We continue to give prompts within the same conversation. The prompt used here is “Walk me through the Transient Analysis on LTspice.” GPT then provided a step-by-step instruction on transient analysis on LTspice for the exact circuit we constructed before. The steps include circuit setup, simulation command configuration, input voltage setup, and result plotting. As shown in Figure \ref{Transient Analysis Vin Configuration by GPT-4o}, GPT can still follow the circuit specifications and provide corresponding reference configurations for the analysis. We conclude that with appropriate conversions to allow GPT to understand the current construction and specification of the circuit, GPT has the capability to provide a good enough instruction on how to conduct the analysis on the certain simulation tool with reasonable reference configurations. \\

GPT also suggested adding a load current step to test the circuit’s ability of handling load transients. As shown in Figure \ref{Optional Analysis Step by GPT-4o}, we can see this step was labeled “optional.” Labelling the additional step is user-friendly. That means GPT is able to provide extra suggestions without misleading the users to conduct any additional analysis by mistake. \\

By the end of the instruction, GPT provided the expected observations for a successful transient analysis simulation as shown in Figure \ref{Expected Observations by GPT-4o} and some simple correction actions for possible failures in Figure \ref{Fine Tuning Suggestions by GPT-4o}. The descriptions for these are not detailed, but still understandable and can give users a relatively comprehensive understanding of the purposes and principles of the analysis. \\


We conclude that when the circuit itself is functionable and complete, GPT-4o can be capable of guiding analysis and simulation for it. The instructions are reasonable and user-friendly. This part of functionality could be a significant assistant to circuit designers, especially the amateurs, to conduct simulation analysis and figure out potential problems.

\section{Conversation-based structure modification}
\subsection{AD8034 operational amplifier}
\begin{figure}
\centering
\includegraphics[width=0.35\textwidth, height=0.2\textwidth]{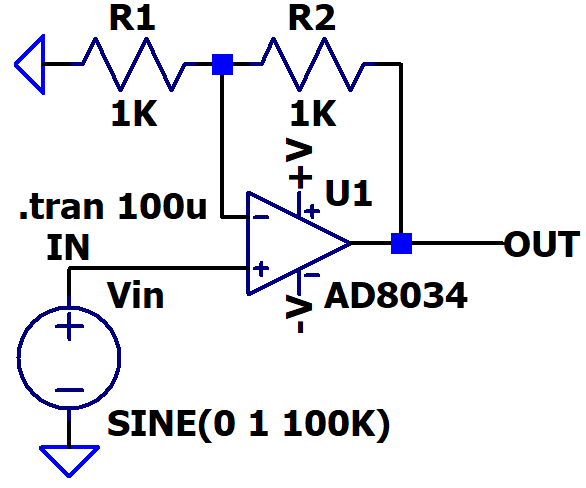}
\caption{AD8034 operational amplifier}
\label{AD8034 operational amplifier}
\end{figure}

Key performance metrics can be measured to evaluate the operational amplifier's functionality (figure \ref{AD8034 operational amplifier}). The output response can be observed on an oscilloscope or multimeter, ensuring Vout follows the expected behavior \cite{xu_2025}. For input-output relationships, apply a step or sinusoidal input and measure the output amplitude and phase.

Using a sinusoidal power supply, such as SINE(Voffset=0, Vamp=1, Freq=100k), for the op-amp's positive terminal creates a test scenario. The behavior observed indicates that the op-amp is functioning in an inverting amplifier configuration and successfully amplifies the input signal. 

\begin{figure}
\centering
\begin{subfigure}{0.23\textwidth}
\centering
\includegraphics[width=\textwidth, height=0.8\textwidth]{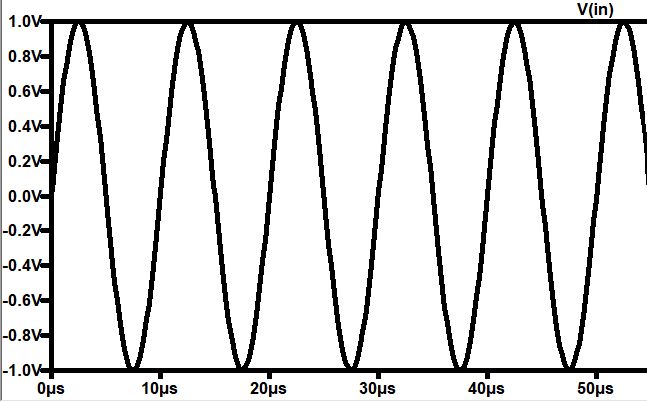}
\caption{input}
\label{AD8034 input sin}
\end{subfigure}
\begin{subfigure}{0.23\textwidth}
\centering
\includegraphics[width=\textwidth, height=0.8\textwidth]{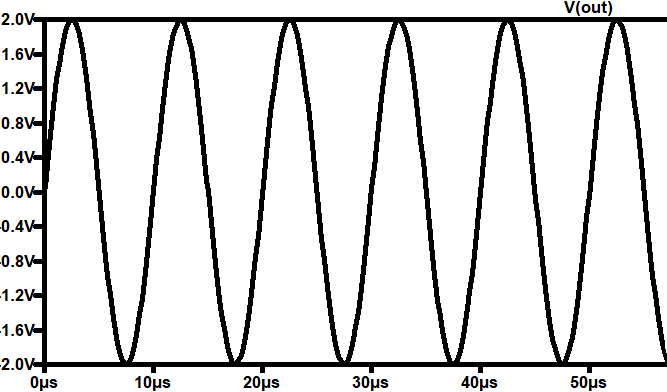}
\caption{output}
\label{AD8034 output sin}
\end{subfigure}
\caption{AD8034 with sinusoidal power supply}
\end{figure}

The input voltage oscillates between -1V and +1V in figure \ref{AD8034 input sin}, indicating that the signal applied to the inverting input is sinusoidal and symmetric about 0V. The output voltage oscillates between -2V and +2V in figure \ref{AD8034 output sin}, demonstrating a gain of \(A_v = -2\). This means the op-amp amplifies the input voltage by a factor of 2 and inverts the signal phase, which is characteristic of the inverting configuration. The output voltage is shown in figure \ref{output voltage, single stage amplifier}.

In an inverting amplifier configuration, the relationship between the input and output is given by \(V_{\text{out}} = -A_v \cdot V_{\text{in}} = -\frac{R_2}{R_1} \cdot V_{\text{in}}\). From the observations, \(A_v = -2\), which indicates that \(\frac{R_2}{R_1} = 2\). This is consistent with the resistor ratio used in the feedback loop and the input resistor.

Although op-amps are typically powered by DC supplies, the sinusoidal power supply effectively fluctuates the operating range of the op-amp between +1V and -1V. The input signal (\(-1 \leq V_{\text{in}} \leq +1\)) and the amplified output (\(-2 \leq V_{\text{out}} \leq +2\)) remain within the supply limits \cite{lai_2021}. This dynamic operation demonstrates that the op-amp can amplify without clipping, as the input and output voltages are within the instantaneous voltage provided by the power supply.

This behavior confirms that the op-amp responds to the input and produces a clean, amplified output consistent with the gain \(A_v = -2\). This indicates that the internal transistors and feedback mechanisms of the op-amp are functioning properly despite the unusual power supply \cite{hersh_2001}. Replacing the sinusoidal power supply with a stable DC voltage would allow evaluation of the op-amp over its full operating range. Further tests can measure parameters such as offset voltage, slew rate, bandwidth, and PSRR, which provide a more complete assessment of performance.
\begin{figure}
\centering
\includegraphics[width = 0.4\textwidth, height = 0.35\textwidth]{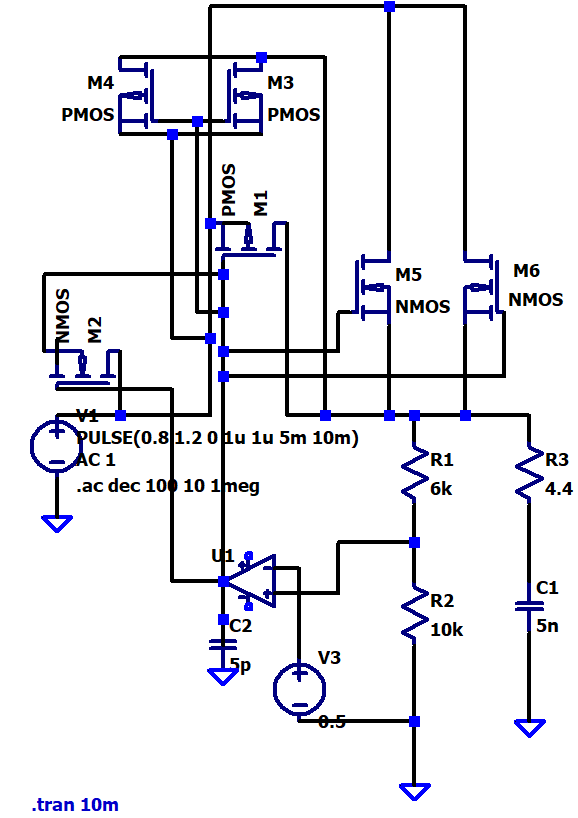}
\caption{LDO AD8034 with 6 transistors}
\label{ldo ad8034 with 6 transistors}
\end{figure}
\begin{figure}
\centering
\includegraphics[width = 0.35\textwidth, height = 0.23\textwidth]{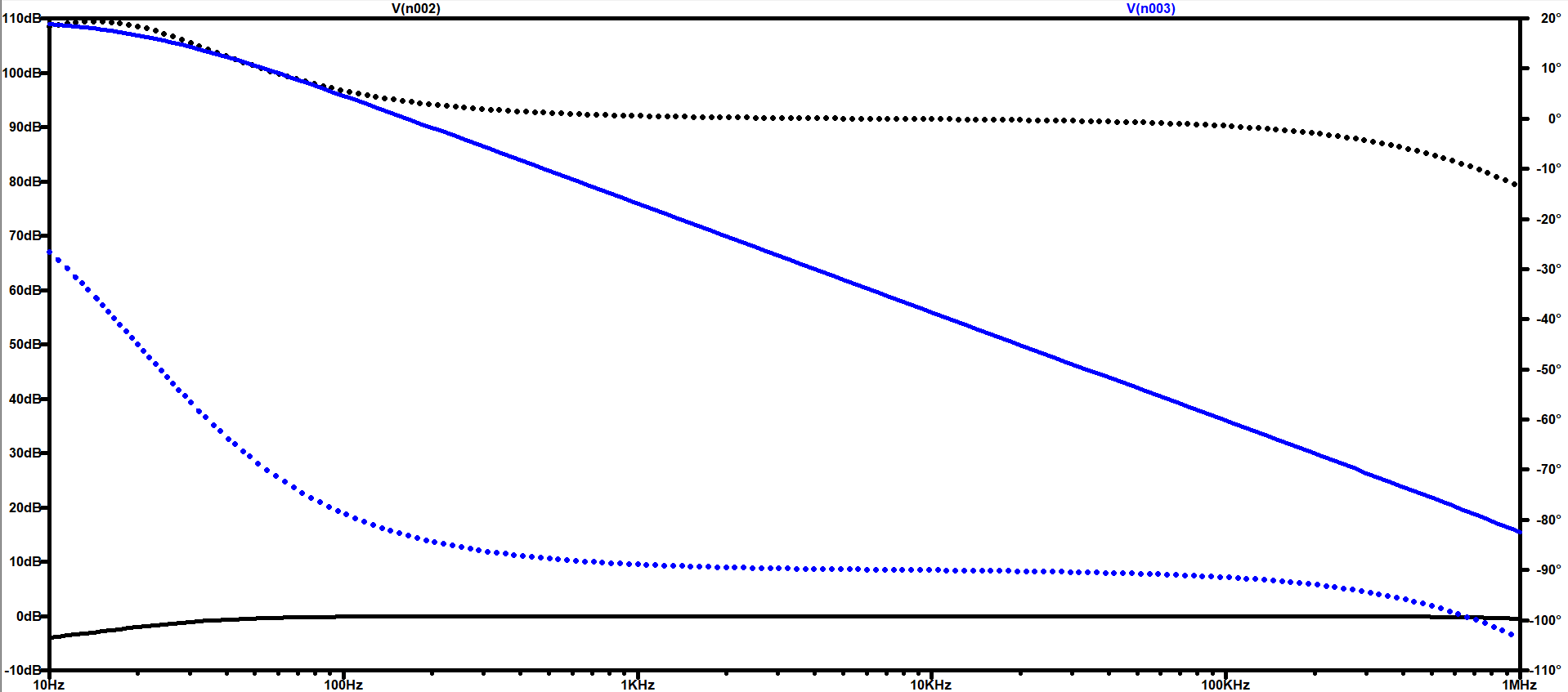}
\caption{frequency response graph, LDO AD8034 with 6 transistors}
\label{frequency response graph, ldo ad8034 with 6 transistors}
\end{figure}

\subsection{Single stage operational amplifier}
\begin{figure}
\begin{subfigure}{0.23\textwidth}
\centering
\includegraphics[width = \textwidth, height = 0.8\textwidth]{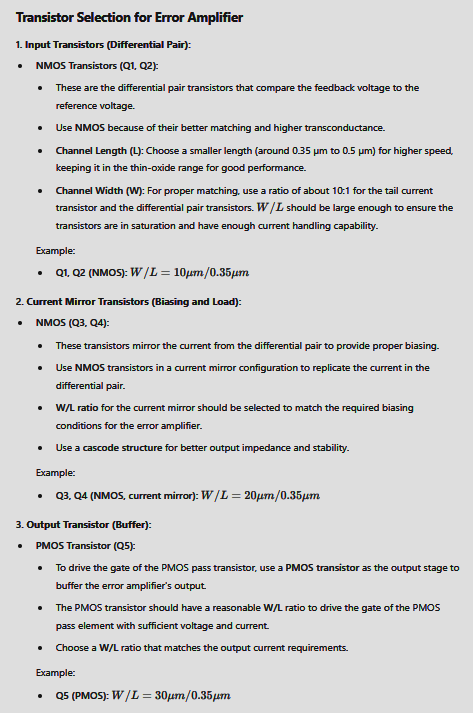}
\caption{transistor selection for single stage operational amplifier}
\label{transistorselection}
\end{subfigure}
\begin{subfigure}{0.23\textwidth}
\centering
\includegraphics[width=\textwidth, height=0.8\textwidth]{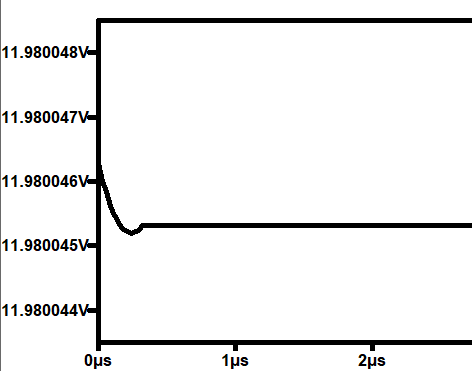}
\caption{voltage output for LDO with ad8034 amplifier}
\end{subfigure}
\caption{transistor characteristics}
\end{figure}

The following prompt is used: ``modify the circuit schematic to improve the operational amplifier". ChatGPT suggested a single-stage operational amplifier with several transistors (figure \ref{transistorselection}).

\begin{figure}
\centering
\includegraphics[width = 0.3\textwidth, height = 0.3\textwidth]{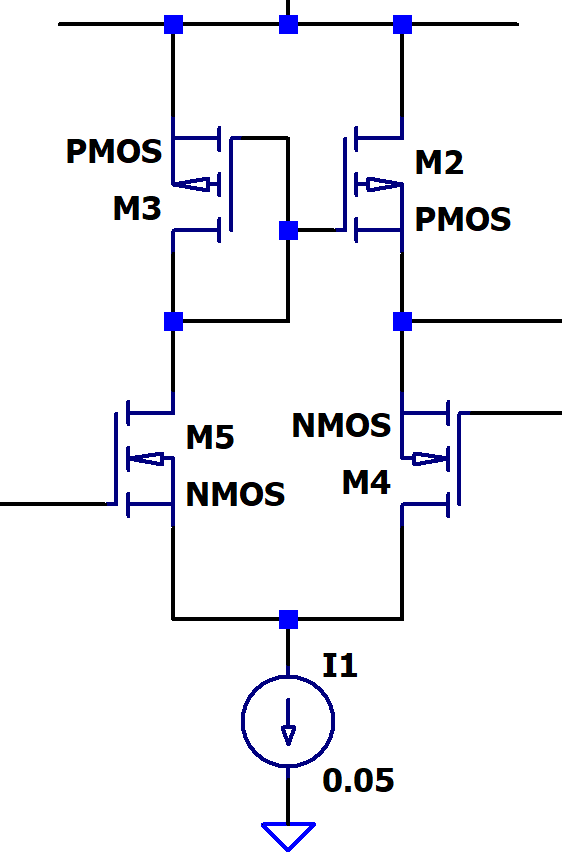}
\caption{single stage operational amplifier}
\label{single stage operational amplifier}
\end{figure}

For the error amplifier in the low-dropout (LDO) regulator design, the transistors should be chosen for their high-speed characteristics and integration capability \cite{sara_2016}. Thin-oxide devices are preferred for their performance. The error amplifier consists of a differential pair, a current mirror, and a buffer stage (figure \ref{single stage operational amplifier}).
\begin{figure}
\centering
\includegraphics[width = 0.5\textwidth, height = 0.4\textwidth]{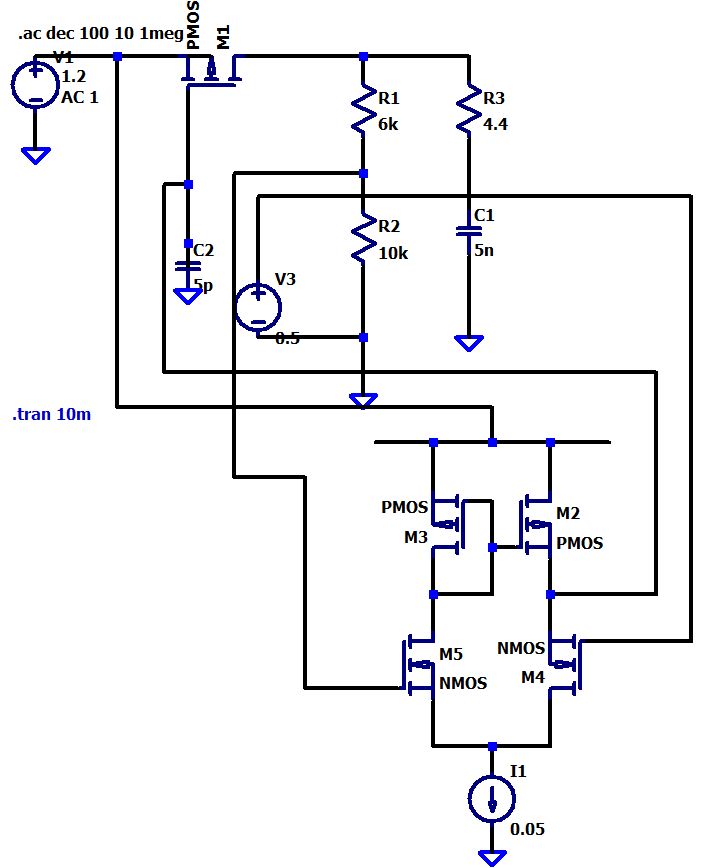}
\caption{LDO with single stage amplifier}
\label{LDO with single stage amplifier}
\end{figure}
\begin{figure}
\centering
\includegraphics[width = 0.35\textwidth, height = 0.24\textwidth]{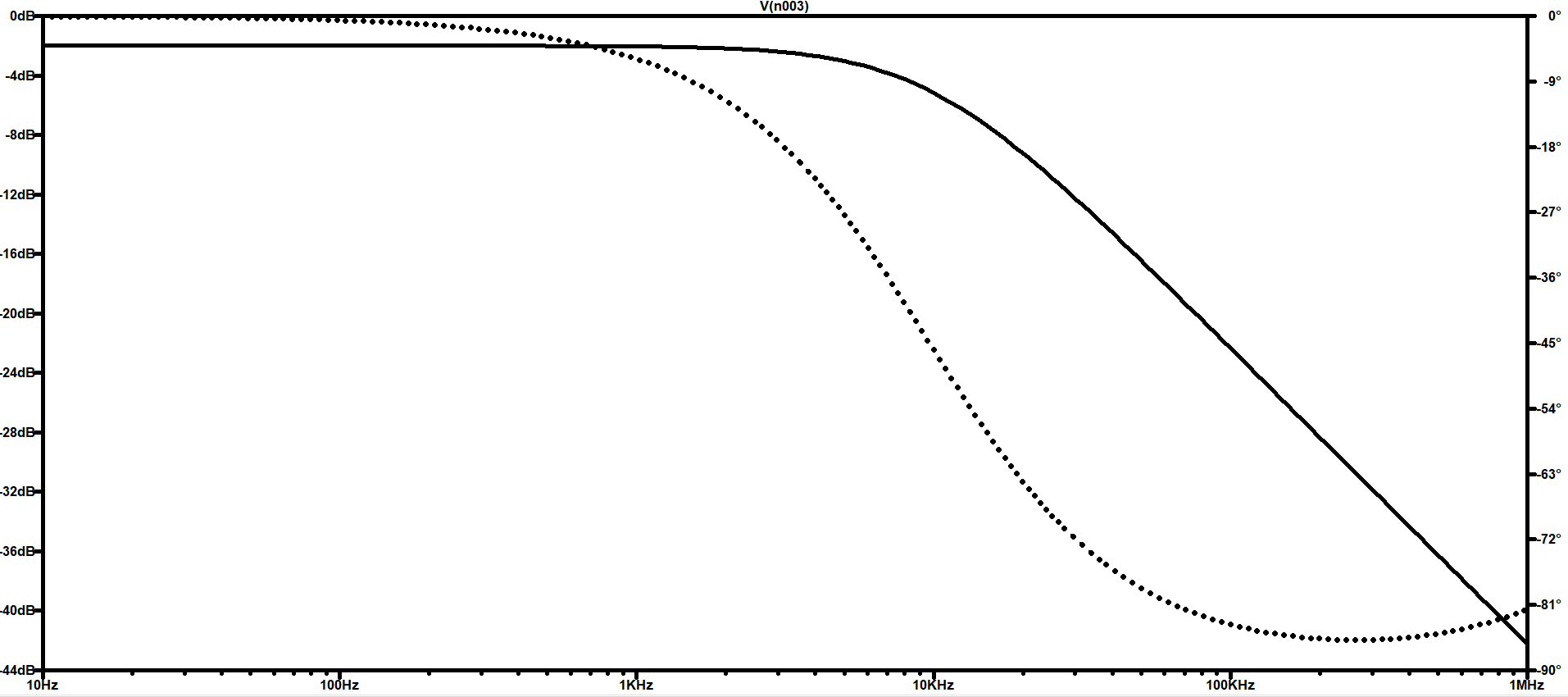}
\caption{frequency response graph, LDO with single stage amplifier}
\label{frequency response graph, single stage amplifier}
\end{figure}

The input transistors form the differential pair and are typically NMOS transistors because of their better matching and higher transconductance \cite{leung_1999}. These transistors compare the feedback voltage to the reference voltage. For these NMOS transistors, a channel length \( L \) of around 0.35 µm to 0.5 µm is suitable for higher speed, keeping the devices in the thin-oxide range for good performance (figure \ref{LDO with single stage amplifier}). The channel width \( W \) should be selected so that the differential pair transistors are in saturation, with a \( W/L \) ratio of around 10:1 for proper matching \cite{tov_2004}. A typical choice could be \( W/L = 10 \, \mu m / 0.35 \, \mu m \) for the input transistors.

The current mirror transistors are used for biasing and providing the necessary current to the differential pair. These transistors should also be NMOS and configured in a current mirror. The NMOS transistors in the current mirror replicate the current from the differential pair. A cascode structure is often used for these transistors to increase output impedance and improve performance \cite{lichen_2024}. The channel width \( W \) and length \( L \) for the current mirror should ensure proper biasing of the error amplifier. A suitable choice could be \( W/L = 20 \, \mu m / 0.35 \, \mu m \).

To drive the gate of the PMOS pass transistor, an output buffer stage is required. This stage is typically implemented using a PMOS transistor, which will buffer the output of the error amplifier. The PMOS transistor needs to have a sufficient \( W/L \) ratio to drive the gate of the pass element effectively. A typical choice for the output transistor could be \( W/L = 30 \, \mu m / 0.35 \, \mu m \).

Optionally, a cascode transistor can be added to the current mirror and the output buffer to improve the overall gain and output impedance. The choice of cascode transistors will depend on the specific design requirements.

The NMOS transistors should have a low threshold voltage (\( V_{th} \)), ideally in the range of 0.2-0.4 V, to ensure that the error amplifier operates correctly even at low input voltages. The error amplifier should be designed to operate with the supply voltage provided to the LDO, typically around 1.2 V.

If the power supply voltage is reduced, the available gate drive for the PMOS pass device diminishes. To sustain acceptable output current levels, the aspect ratio of the power transistor must be increased. However, this adjustment also raises the parasitic gate capacitance. As the size of the PMOS transistor grows, the parasitic capacitance (\(C_{\text{par}}\)) increases, which shifts the parasitic pole (\(P_3\)) to lower frequencies. Consequently, the phase margin of the system decreases, potentially jeopardizing stability, particularly in environments with low quiescent current.

A technique to improve the gate drive without increasing the input voltage or enlarging the device size is to forward-bias the source-to-bulk junction of the PMOS pass device. This method lowers the threshold voltage, a phenomenon referred to as the body effect. The threshold voltage (\(V_{\text{th}}\)) is described by
\[
|V_{\text{th}}| = |V_{\text{to}}| + \gamma \left(\sqrt{2\phi_f - V_{\text{sb}}} - \sqrt{2\phi_f}\right)
\]
where \(V_{\text{to}}\) is the zero-bias threshold voltage, \(\gamma\) is the body effect coefficient, 
\(\phi_f\) is the Fermi potential, and \(V_{\text{sb}}\) is the source-to-bulk voltage.

The maximum output current for comparative analysis is observed in the region where the power PMOS device operates in saturation, which corresponds to the non-dropout condition. The drain current (\(I_{sd}\)) of the PMOS device is
\begin{align*}
&I_{sd} \approx K_p \frac{W}{L} \left( V_{sg} - V_{th} \right)^2\\
&\approx K_p \frac{W}{L} 
\left( V_{sg} - \sqrt{V_{to} - \gamma \left( \sqrt{\phi_f + V_{sb}} - \sqrt{\phi_f} \right)} \right)^2
\end{align*}
where \(K_p\) is the transconductance parameter of the PMOS transistor. 
The maximum output current is achieved when the gate drive reaches its maximum value, 
which occurs when the source-to-gate voltage (\(V_{sg}\)) equals the input voltage (\(V_{in}\)).

The LDO employs a classical two-stage architecture: PMOS pass transistor M1, regulated by a CMOS differential operational amplifier constructed from transistors M3, M2 (PMOS current mirror load), M4, M5 (NMOS differential input pair), and bias current source I1 = $50\,\mu\text{A}$. Fundamental metal-oxide-semiconductor field-effect transistor (MOSFET) physics governs the amplifier small-signal behaviour, open-loop gain, frequency response, and closed-loop stability of the regulator.

For long-channel MOSFETs operating in saturation, the drain current is
\begin{equation*}
I_D = \frac{1}{2}\mu C_{\text{ox}} \frac{W}{L}\left(V_{\text{gs}}-V_{\text{th}}\right)^2 \left(1+\lambda V_{\text{ds}}\right),
\end{equation*}
where $\mu$ denotes carrier mobility, $C_{\text{ox}}$ is gate oxide capacitance per unit area, $W/L$ is transistor aspect ratio, $V_{\text{th}}$ threshold voltage, and $\lambda$ is channel-length modulation parameter. Channel-length modulation introduces finite output resistance for every saturated transistor:
$r_o = \frac{1}{\lambda I_D}$.
This parasitic output resistance dominates the intrinsic gain of the differential amplifier and sets fundamental limits on DC loop gain.

The core amplifier is formed by an NMOS differential pair (M4, M5) biased by tail current source I1. Under differential input voltage $v_{\text{id}}$, the differential pair generates a differential drain current:
\begin{equation*}
i_d = g_m v_{\text{id}},\qquad g_m = \sqrt{2\mu C_{\text{ox}}\frac{W}{L}I_{D,\text{tail}}/2}
\end{equation*}
where $g_m$ is the transconductance of the input transistors. The PMOS devices M2 and M3 implement a current-mirror active load, converting differential drain current into single-ended output voltage driving the gate of PMOS pass transistor M1. The unloaded DC voltage gain of the differential amplifier stage equals
\begin{equation*}
A_{v,\text{diff}} = -g_{m4} \left(r_{o4}\parallel r_{o2}\right).
\end{equation*}
The parallel combination $(r_{o4}\parallel r_{o2})$ is the effective load resistance of the differential amplifier. Any reduction in transistor output resistance lowers DC gain and degrades low-frequency regulation precision.

The PMOS pass element M1 operates as a common-source power transistor within the regulation loop. Its small-signal transconductance $g_{m,\text{M1}}$ creates an additional gain stage. The composite open-loop gain of the full regulator becomes
\begin{equation*}
A_{\text{OL,DC}} = A_{v,\text{diff}} \cdot g_{m,\text{M1}} \cdot R_{\text{load,eff}} \cdot \beta
\end{equation*}
with feedback factor $\beta = R_2/(R_1+R_2)$ defined by the resistive divider $R_1$, $R_2$. The reference voltage $V_3=0.5\,\text{V}$ sets the DC equilibrium condition $V_{\text{sense}} = V_3$.

Parasitic capacitances impose frequency rolloff and determine the loop phase response, critical for stability analysis performed via the simulation command \texttt{.ac dec 100 10 1meg}. Gate-drain overlap capacitance $C_{\text{gd}}$ and gate-source capacitance $C_{\text{gs}}$ exist on all MOS devices. Gate capacitance $C2=5\,\text{pF}$ at the gate node of M1 acts as an explicit compensation capacitor. This capacitance introduces the dominant pole of the open-loop transfer function:
$\omega_p \approx \frac{1}{R_o C_2}$,
where $R_o$ is the amplifier output resistance driving M1’s gate. Additional non-dominant poles emerge from the output load network formed by $R_3=4.4\,\Omega$ and $C_1=5\,\text{nF}$. At high frequencies, these poles add phase lag and erode phase margin.

Transient behaviour, evaluated using \texttt{.tran 10m}, is governed by charge dynamics across all transistor capacitances. When the LDO experiences load perturbations, the amplifier must adjust the gate voltage of M1 to stabilise the output voltage. The rate of gate voltage change is limited by the available charging/discharging current from the differential amplifier and the compensation capacitance $C2$. Slow gate response creates voltage undershoot/overshoot, while insufficient phase margin leads to sustained underdamped transient ringing.

All capacitive and resistive device parasitics couple with external passive components $R_1,R_2,R_3,C_1$ to shape the regulator’s small-signal transfer function (figure \ref{ldo ad8034 with 6 transistors}). When additional inductive magnetic elements are inserted into the circuit, their frequency-dependent impedance introduces supplementary phase shift. If such elements reside inside the feedback path, they modify the effective feedback factor $\beta(s)$ and interact with the inherent frequency limitations of the CMOS amplifier to further compromise stability. In contrast, inductors placed after the feedback sensing tap only alter the load impedance and do not directly modify the closed-loop amplifier dynamics.

Using the four transistor single stage amplifier, the frequency response graph shown at figures \ref{frequency response graph, single stage amplifier}, \ref{frequency response graph, ldo ad8034 with 6 transistors} likely represent the gain and phase behavior of the LDO regulator's control loop with a single-stage operational amplifier. The two curves in the graph correspond to the magnitude (gain) and phase of the open-loop transfer function.

The solid line shows the magnitude of the open-loop gain of the LDO in decibels (dB) as a function of frequency. At low frequencies, from 10 Hz to approximately 1 kHz, the gain remains relatively constant and high, representing the DC gain of the single-stage amplifier. This high gain is essential for accurate regulation and minimizing output voltage error. As frequency increases, the gain begins to decrease at a certain rate due to the presence of poles in the circuit, typically originating from the output capacitor, compensation network, and internal parasitic capacitances. The slope of the gain curve is approximately \( -20 \, \text{dB/decade} \), characteristic of a dominant pole system. This indicates that the single-stage amplifier has a well-defined dominant pole responsible for stability and loop response.

The dashed line shows the phase shift in the control loop as a function of frequency. At low frequencies, the phase is close to \( 0^\circ \), indicating minimal phase shift in the loop. As frequency increases, the phase begins to drop, eventually approaching \( -90^\circ \). This phase lag is due to the presence of poles in the transfer function. The dominant pole introduces a \( -90^\circ \) phase shift, which stabilizes the system while limiting the bandwidth. The phase asymptotically approaches \( -90^\circ \) and does not reach \( -180^\circ \), indicating that the circuit remains stable within the frequency range shown.

The bandwidth of the system is defined by the frequency at which the gain falls to unity (0 dB), referred to as the unity-gain bandwidth. This bandwidth determines how quickly the LDO can respond to changes in load current or input voltage. Stability is determined by the phase margin, which is the difference between \( -180^\circ \) and the phase at the unity-gain frequency. A phase margin of at least \( 45^\circ \) is typically required for stable operation without oscillations. 

This graph demonstrates that the LDO regulator with a single-stage amplifier has a stable frequency response with adequate gain and phase characteristics for maintaining good regulation performance. The use of compensation, such as the compensation capacitor \( C_c \), and the small output capacitor ensures the stability of the system while providing sufficient bandwidth.

\subsection{Adding inductors to the LDO circuit topologies}
\begin{figure}
\centering
\begin{subfigure}{0.23\textwidth}
    \centering
    \includegraphics[width=\textwidth,height=0.8\textwidth]{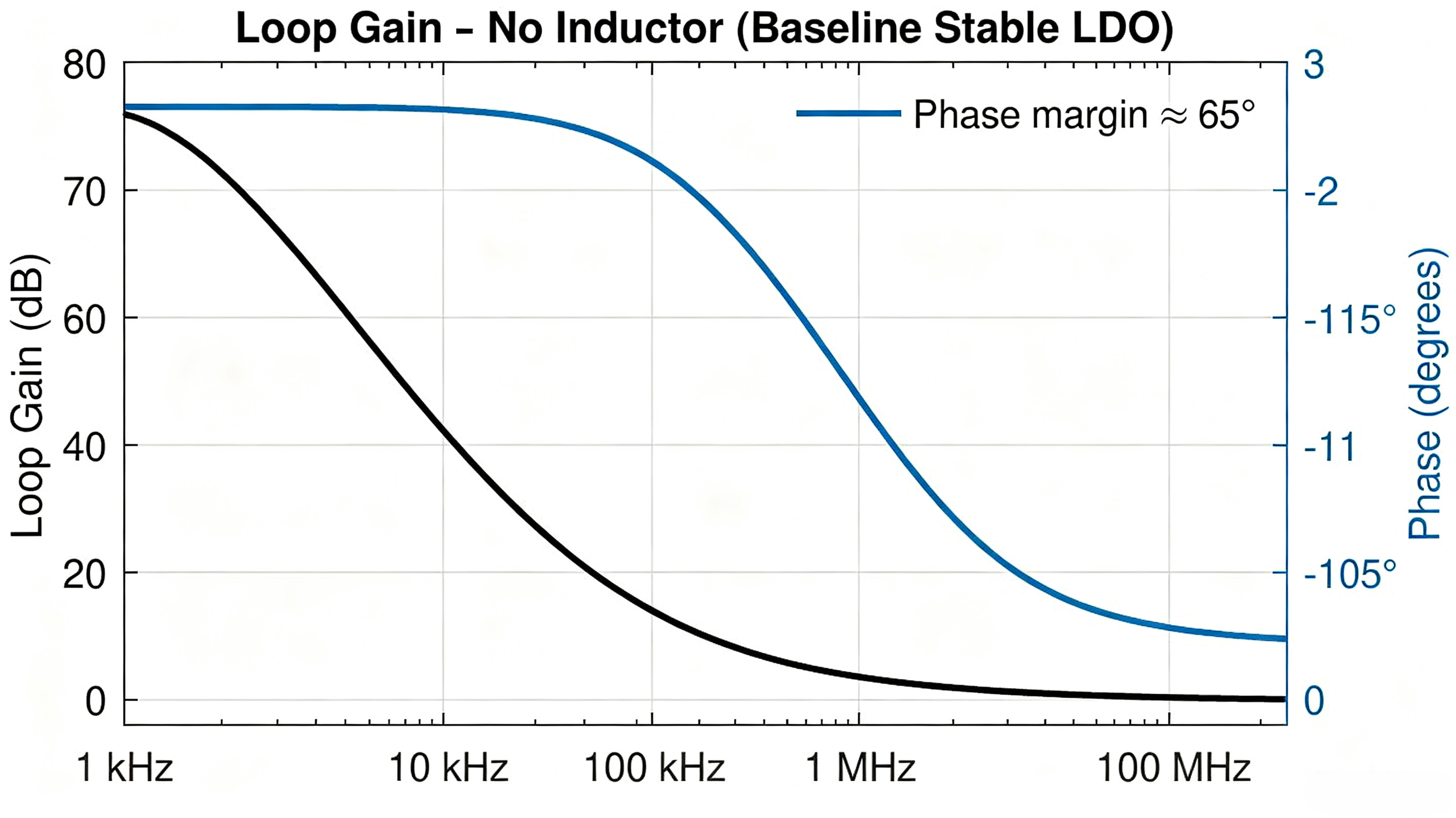}
    \caption{Loop gain response, baseline LDO (no inductor). Stable single-pole rolloff with sufficient phase margin.}
    \label{fig:bode_baseline}
\end{subfigure}
\begin{subfigure}{0.23\textwidth}
    \centering
    \includegraphics[width=\textwidth,height=0.8\textwidth]{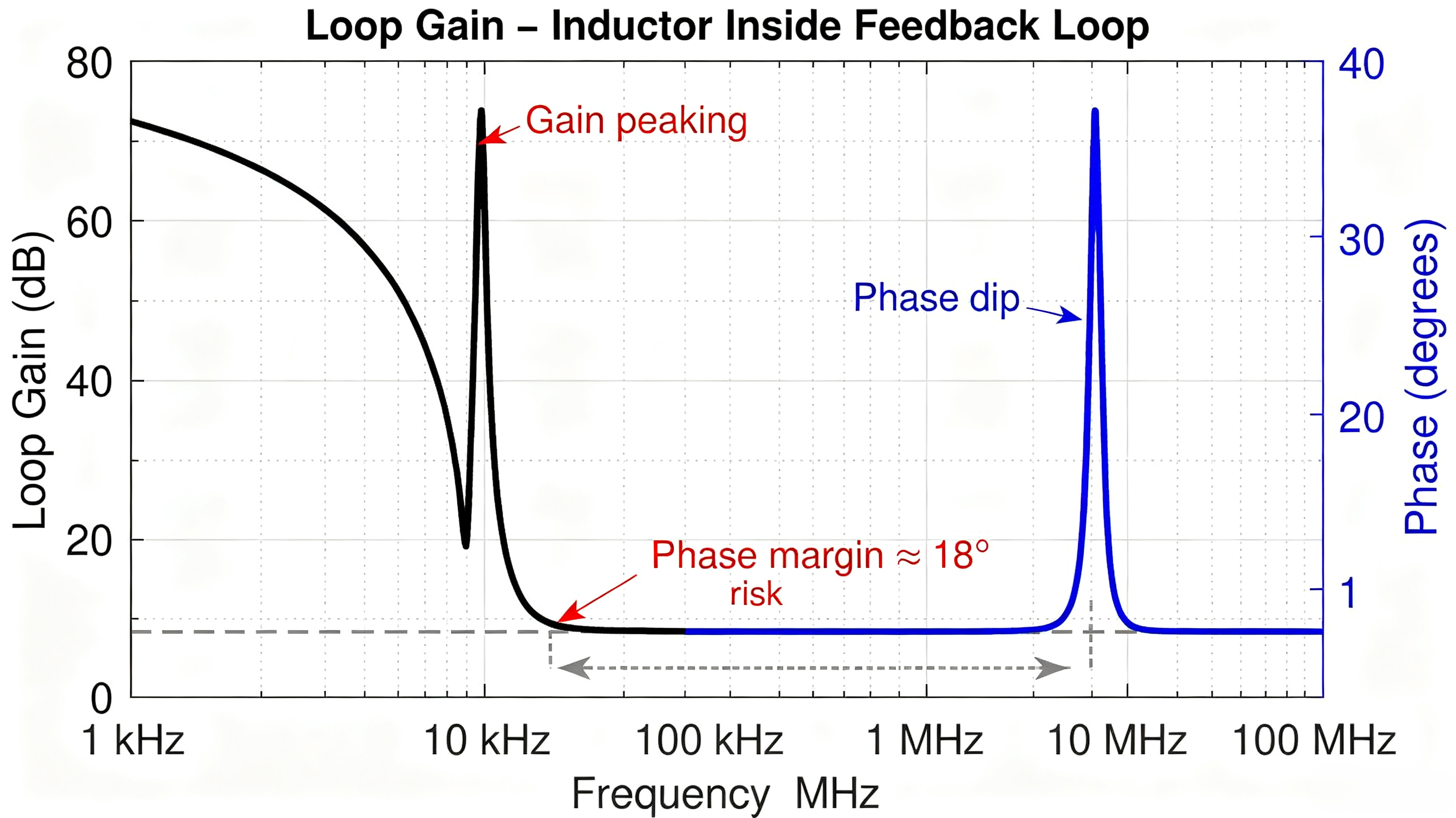}
    \caption{Loop gain response, inductor placed within feedback path. Gain peaking and sharp phase drop arise from LC poles.}
    \label{fig:bode_inside_loop}
\end{subfigure}
\begin{subfigure}{0.23\textwidth}
    \centering
    \includegraphics[width=\textwidth,height=0.8\textwidth]{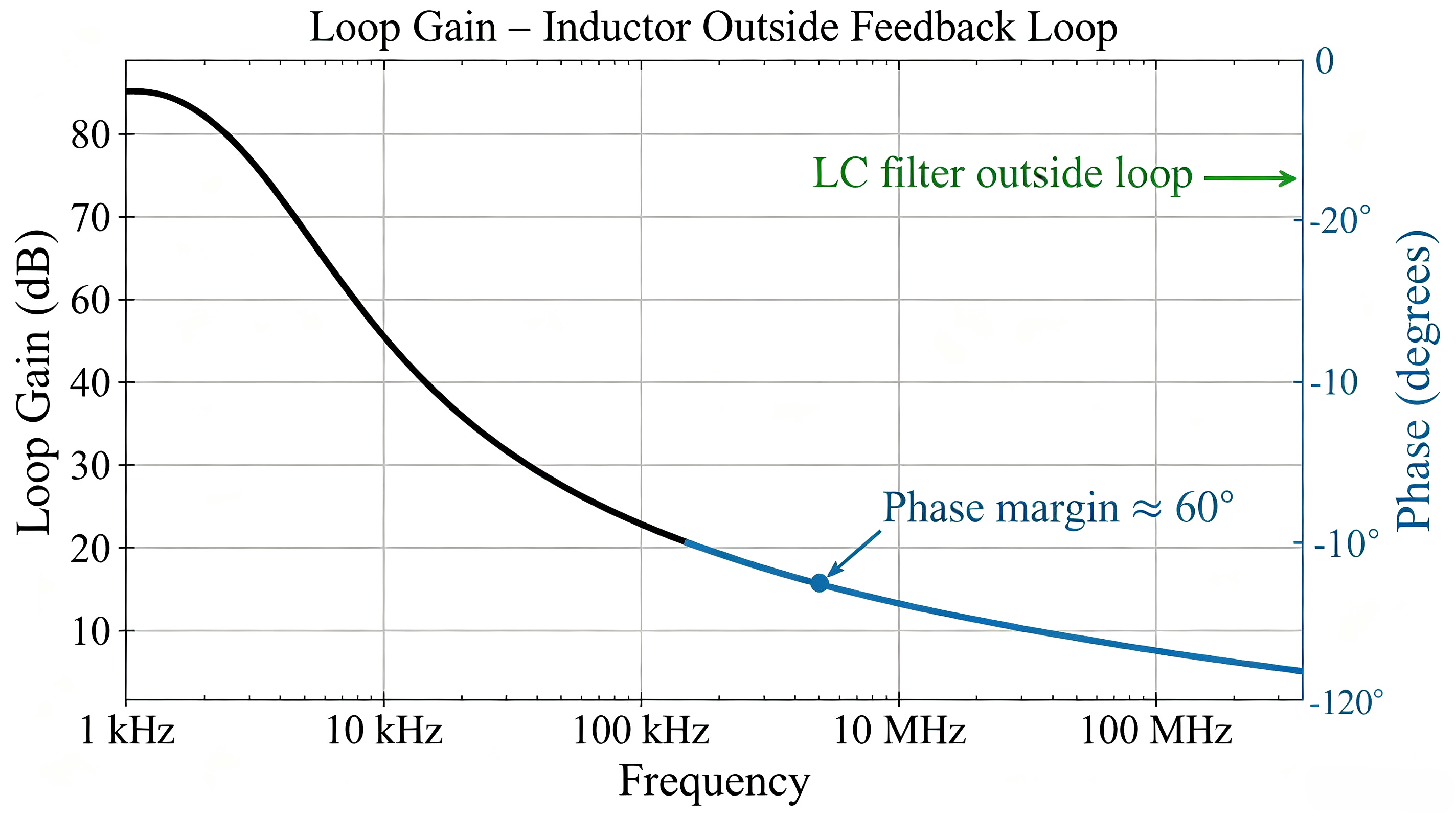}
    \caption{Loop gain response, inductor placed outside feedback path. The LC filter lies outside the control loop.}
    \label{fig:bode_outside_loop}
\end{subfigure}
\begin{subfigure}{0.23\textwidth}
    \centering
    \includegraphics[width=\textwidth,height=0.8\textwidth]{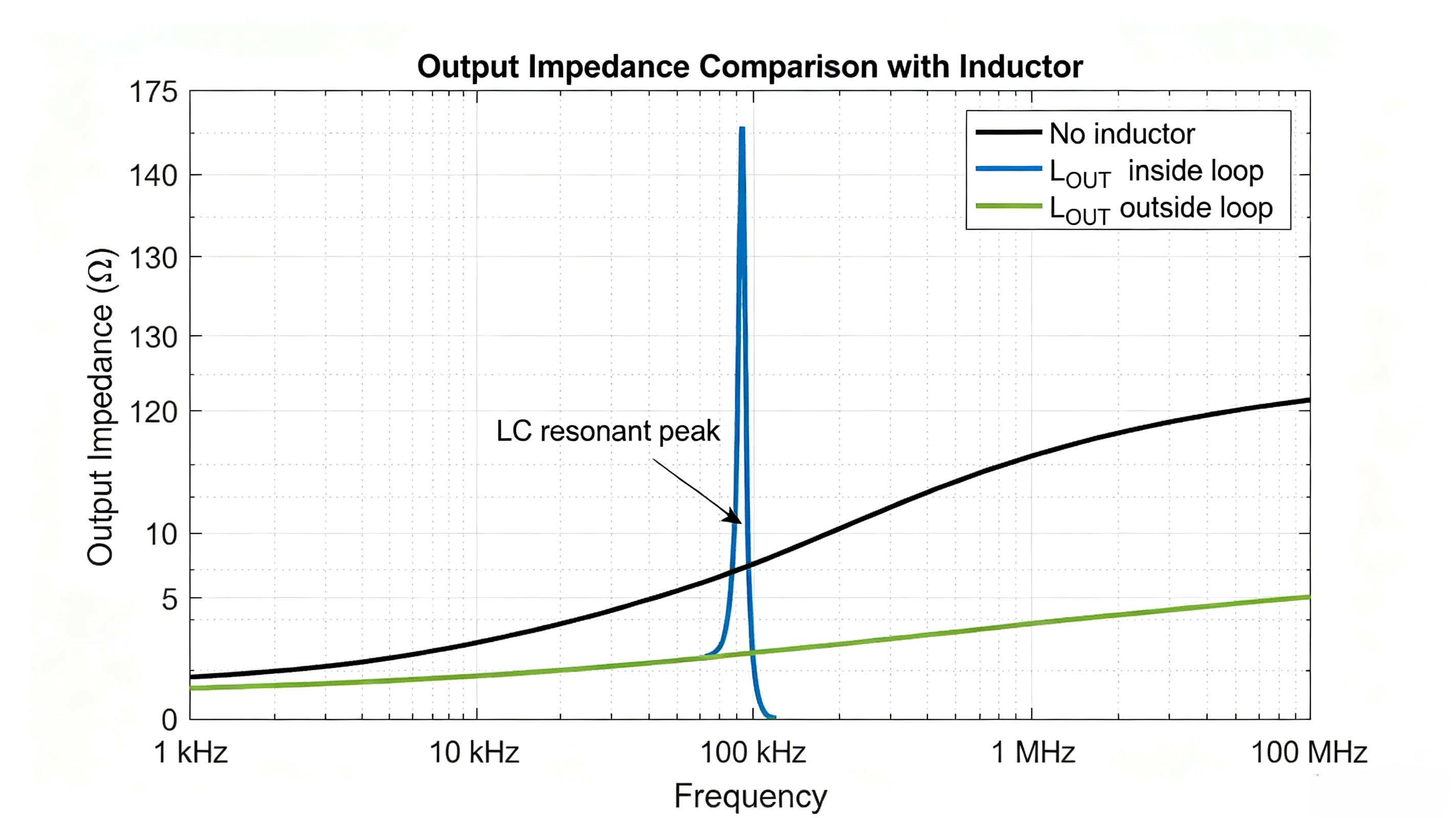}
    \caption{Output impedance frequency compare, resonant impedance peak if LC in regulator feedback loop.}
    \label{fig:zout_compare}
\end{subfigure}
\caption{AC small-signal analysis for the LDO topology under different inductor placement configurations.}
\label{fig:ldo_ac_comparison}
\end{figure}
\begin{figure}
\centering
\includegraphics[width=0.45\textwidth,height=0.34\textwidth]{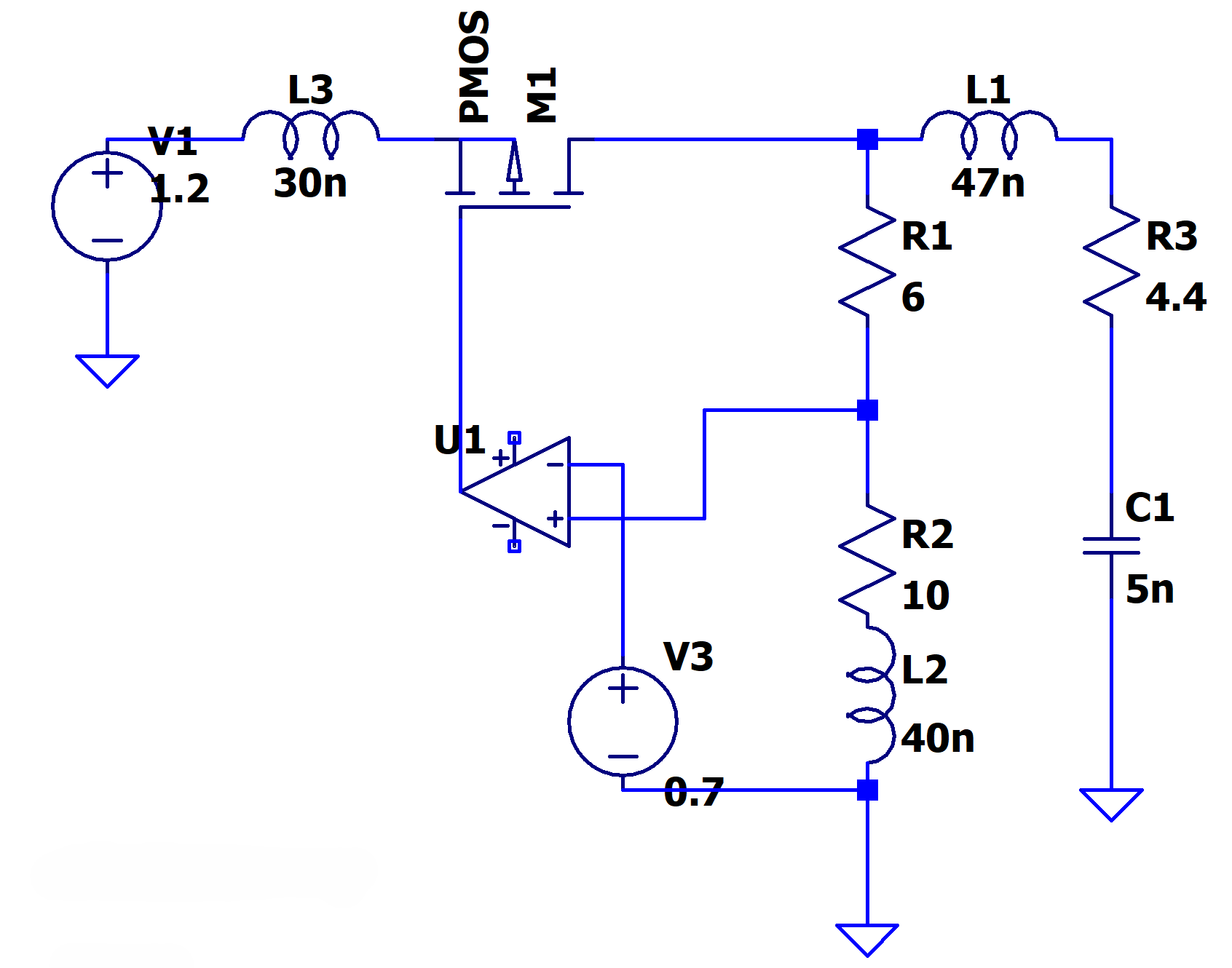}
\caption{LDO circuit topology with multiple inductors. $L_3=30nH$ provides input-side supply noise filtering, $L_1=47nH$ forms an $LC$ output filter with $C_1=5nF$ after feedback sensing node, and $L_2=40nH$ is inserted in the lower leg of feedback resistor divider. PMOS transistor M1 acts as the pass device, and opamp U1 implements closed-loop voltage regulation referenced to $V_3=0.7\mathrm{V}$.}
\label{fig:ldo_schematic_inductor_variants}
\end{figure}
The presented circuit implements a classic voltage-mode low-dropout regulator (LDO) constructed using an operational amplifier and a PMOS pass transistor. Inductors may be incorporated at three distinct locations within the topology, with drastically different stability implications and functional purposes \cite{gruosso_2019}. Critically, inductors within a linear LDO act only as AC filtering elements and cannot improve power conversion efficiency, unlike inductors in switching converters; inductors behave as DC short circuits under steady-state operating conditions and produce no shift to the DC regulated output voltage.

All inductive components within the LDO topology are electromagnetic systems governed by the macroscopic Maxwell equations in differential form
\begin{align*}
&\nabla\times\mathbf{E} = -\frac{\partial\mathbf{B}}{\partial t},\quad
\nabla\cdot\mathbf{D} = \rho_v\\
&\nabla\times\mathbf{H} = \mathbf{J} + \frac{\partial\mathbf{D}}{\partial t},\quad
\nabla\cdot\mathbf{B} = 0
\end{align*}
Faraday’s law of induction establishes the physical origin of inductive voltage. Integrating over a closed conductive contour $\mathcal{C}$ bounding surface $\mathcal{S}$ yields the integral form
\begin{equation*}
\oint_{\mathcal{C}} \mathbf{E}\cdot d\mathbf{l} = -\frac{d}{dt}\iint_{\mathcal{S}} \mathbf{B}\cdot d\mathbf{S} = -\frac{d\Psi}{dt},
\end{equation*}
where $\Psi = \iint_{\mathcal{S}} \mathbf{B}\cdot d\mathbf{S}$ denotes the magnetic flux linked to the circuit contour. For a lumped inductor with $N$ turns, the flux linkage is defined as $\Lambda = N\Psi$ \cite{LIU2024}. The terminal voltage across the inductor is $v_L = \frac{d\Lambda}{dt}$.
Linear inductance $L$ is defined as the proportionality constant between flux linkage and coil current $i$ as
$\Lambda = Li,\quad v_L = L\frac{di}{dt}$,
which is the lumped-circuit approximation derived directly from Faraday’s law, valid when device physical dimensions are much smaller than the electromagnetic wavelength of signal frequencies \cite{HURLEY2018}. Taking the Laplace transform under zero initial conditions recovers the frequency-domain impedance:
$V_L(s) = sLI_L(s)\quad Z_L(s) = sL$.

Ampère’s circuital law governs the magnetic field generation by current flowing through conductors in the LDO. At low-to-moderate frequencies, the displacement current term $\partial\mathbf{D}/\partial t$ can be neglected for magnetic field calculation ($\nabla\times\mathbf{H}\approx\mathbf{J}$), forming the basis of quasi-static magnetostatics used to estimate inductance values for $L_1$, $L_2$ and $L_3$.

For homogeneous, linear, isotropic materials, the constitutive relations close the system
\begin{equation*}
\mathbf{B} = \mu\mathbf{H},\quad \mathbf{D}=\varepsilon\mathbf{E},\quad \mathbf{J}=\sigma\mathbf{E}
\end{equation*}
Finite conductivity $\sigma$ introduces series parasitic resistance (DCR) in physical inductors, while permittivity $\varepsilon$ models inter-turn parasitic capacitance responsible for self-resonance at high frequencies.

Within the LDO circuit shown in Figure \ref{fig:ldo_schematic_inductor_variants}, three inductors represent spatially separated electromagnetic systems: input filter $L_3$, output load filter $L_1$, and feedback-divider inductor $L_2$. Each inductor obeys the Faraday-law derived voltage-current relation. Critically, the topological placement of each electromagnetic element determines whether its dynamic impedance enters the regulator closed-loop transfer function.
When an inductor is embedded inside the feedback signal path (such as $L_2$), its frequency-dependent impedance modifies the feedback factor $\beta(s)$ and introduces phase lag originating from time-varying magnetic flux. This electromagnetic phase shift propagates around the control loop, reduces phase margin, and generates underdamped voltage oscillations observed in transient simulations. In contrast, inductors placed outside the feedback sensing tap (such as $L_1$ and $L_3$) produce flux-induced voltage dynamics that only affect the load or input supply, without perturbing the core feedback signal.

This first-principles framework confirms that all observed stability behaviour originates from the electromagnetic induction captured by Maxwell’s equations, and justifies the use of lumped $sL$ impedance models for stability analysis of the LDO with integrated magnetic circuit elements.

All inductive components are modeled using linear frequency-domain Laplace representation, where the impedance of an inductor is
\begin{equation*}
Z_L(s) = sL = \mathrm{j}\omega L
\end{equation*}
neglecting parasitic series resistance and self-resonance for the base analysis. The core closed-loop regulator transfer function of the LDO without additional inductors is
\begin{equation*}
A_{\mathrm{OL}}(s) = A_{\mathrm{amp}}(s) \cdot A_{\mathrm{M1}}(s)\cdot\beta_0
\end{equation*}
where $A_{\mathrm{amp}}(s)$ is the op-amp gain, $A_{\mathrm{M1}}(s)$ denotes the small-signal gain of the PMOS pass device, and $\beta_0$ is the ideal resistive feedback factor:
\begin{equation*}
\beta_0 = \frac{R_2}{R_1+R_2}
\end{equation*}
Three distinct configurations for magnetic inductive elements are analysed in the circuit shown in Figure \ref{fig:ldo_schematic_inductor_variants}.
The input-side inductor $L_3$ lies between the supply source and the source terminal of PMOS M1. For small-signal perturbations, the supply node is treated as an AC ground, so $L_3$ only modifies supply-domain noise filtering and does not enter the control loop transfer function \cite{Leupold1997}. Consequently, $L_3$ exerts negligible influence on loop stability.
The output filter inductor $L_1-R_3-C_1$ (outside feedback tap) second-order network is connected after the feedback sensing node. The feedback voltage remains sampled directly at the drain terminal of M1, so the feedback factor $\beta=\beta_0$ remains unchanged. The load network creates an independent output impedance characteristic but is not embedded within the closed-loop signal path. The load transfer function from the regulated node to the load terminal is
\begin{equation*}
H_{\mathrm{load}}(s) = \frac{V_{\mathrm{load}}(s)}{V_{\mathrm{reg}}(s)}
= \frac{1}{s^2 L_1 C_1 + s R_3 C_1 + 1}.
\end{equation*}
With $L_1=47\,\mathrm{nH}$, $C_1=5\,\mathrm{nF}$, $R_3=4.4\,\Omega$, the circuit achieves near-critical damping, minimising load-side voltage ringing while preserving the original LDO loop stability \cite{norman_2017}.
When feedback-divider inductor $L_2$ (inside feedback path) is inserted in the lower branch of the feedback divider, the feedback network becomes frequency-dependent. The modified feedback factor $\beta(s)$ takes the form
\begin{equation*}
\beta(s) = \frac{R_2 + sL_2}{R_1 + R_2 + sL_2}
\end{equation*}
This feedback network introduces additional frequency-domain phase shift into the open-loop transfer function $A_{\mathrm{OL}}(s)=A_{\mathrm{amp}}(s)A_{\mathrm{M1}}(s)\beta(s)$. At high frequencies, the impedance of $L_2$ rises, altering the feedback magnitude and injecting phase lag \cite{scot_2022}. This phase degradation reduces phase margin, which manifests as underdamped oscillations and voltage ringing observed in transient simulations. For sufficiently large $L_2$, the additional phase shift can drive the system towards conditional instability.

The analysis highlights a fundamental distinction: inductors placed outside the feedback sensing loop only modify the load response and do not compromise regulator stability, whereas any magnetic component embedded within the feedback divider creates a frequency-dependent feedback factor that degrades small-signal stability.

First, a series output inductor placed outside the feedback sensing network forms an LC low-pass filter with the output capacitor. This configuration isolates the LDO control loop from the LC resonant poles, preserving loop phase margin and eliminating oscillation risk \cite{mion_2025}. This topology is used to suppress high-frequency electromagnetic interference and damp fast transient load current spikes. Second, a series output inductor inserted inside the feedback path places the second-order LC resonance within the regulator’s control loop \cite{liu2026}. The complex conjugate pole pair from the LC network degrades phase margin, introducing voltage ringing, gain peaking, and potentially sustained oscillation; additional frequency compensation is mandatory if this topology is required. Third, an input-side series inductor forms an input LC filter together with supply decoupling capacitors, blocking conducted noise propagating between the input supply and the LDO, with no direct impact on output voltage dynamics.

For small-signal analysis, the transfer function of each topology must be derived to evaluate loop stability and filtering performance. Transient simulations characterize voltage overshoot and ringing during load-step events, while AC frequency sweeps quantify noise attenuation and phase margin \cite{lwithwaite_1967}. Damping factor analysis for the LC filter imposes a practical upper bound on inductance value to avoid underdamped resonant behaviour. For the given load resistance and output capacitance, near-critical damping is achieved with an inductance near 47\,nH when the filter is placed outside the feedback loop.

Component sizing for the series LC filter is constrained by the target damping characteristics of the second-order $L$-$R_3$-$C_1$ network, with fixed load resistance $R_3 = 4.4\,\Omega$ and output capacitance $C_1 = 5\mathrm{nF}$. The resonant frequency and damping factor are defined by
\begin{equation*}
f_0 = \frac{1}{2\pi\sqrt{L C_1}}, \qquad \zeta = \frac{R_3}{2}\sqrt{\frac{C_1}{L}}
\end{equation*}
where $\zeta$ denotes the damping ratio. To suppress severe transient ringing, a minimum damping ratio of $\zeta \ge 0.5$ is enforced, which yields an upper bound on allowable inductance $L \le R_3^2 C_1 = 96.8\,\mathrm{nH}$.
Candidate inductance values and their corresponding filter performance are summarised in Table \ref{tab:inductor_sweep}.
\begin{table}
\footnotesize
\centering
\caption{Performance metrics for candidate inductance values with fixed $C_1=5\,\mathrm{nF}$, $R_3=4.4\,\Omega$.}
\label{tab:inductor_sweep}
\begin{tabular}{|c|c|c|c|}\hline
inductance & \begin{tabular}{@{}c@{}}resonance\\frequency $f_0$\end{tabular} & \begin{tabular}{@{}c@{}}damping\\ratio $\zeta$\end{tabular} & behavior\\\hline
22 nH & 480 MHz & 1.04 & \begin{tabular}{@{}c@{}}overdamped,\\minimal ringing\end{tabular}\\\hline
47 nH & 330 MHz & 0.71 & \begin{tabular}{@{}c@{}}near critical\\damping\end{tabular}\\\hline
68 nH & 275 MHz & 0.59 & \begin{tabular}{@{}c@{}}light underdamp,\\mild ringing\end{tabular}\\\hline
91 nH & 238 MHz & 0.51 & \begin{tabular}{@{}c@{}}borderline\\underdamped\end{tabular}\\\hline
$>100$ nH & $<225$ MHz & $<0.5$ & \begin{tabular}{@{}c@{}}strong transient\\ringing\end{tabular}\\\hline
\end{tabular}
\end{table}
An inductance of 47\,nH is selected as the baseline design point, achieving near-critical damping to balance high-frequency noise attenuation and suppression of voltage oscillations \cite{sinan_2004}. Hardware implementation requires a low-DCR high-frequency surface-mount inductor with self-resonant frequency at least three times $f_0$ and saturation current exceeding the maximum load current. If the LC filter is placed within the LDO feedback loop, substantially smaller inductance (10-22\,nH) must be adopted to push resonant poles far above the regulator control bandwidth and mitigate stability degradation.

Figure \ref{fig:ldo_schematic_inductor_variants} presents the op-amp based PMOS LDO test circuit constructed to evaluate the influence of inductive components at three distinct locations. The input supply voltage $V_1=1.2\,\mathrm{V}$ feeds the pass transistor M1 through series inductor $L_3$, which acts as an input-side noise filter. Closed-loop regulation is realised by operational amplifier U1, which compares the feedback voltage from the resistive divider $R_1$-$R_2$ against the reference voltage $V_3=0.7\,\mathrm{V}$ and drives the gate of the PMOS pass element. The output node of M1 connects to an external load filter formed by $L_1=47\,\mathrm{nH}$ and $C_1=5\,\mathrm{nF}$ in series with load resistance $R_3=4.4\,\Omega$; this $LC$ filter is positioned after the feedback sensing tap, isolating the resonant network from the control loop to preserve stability. An additional inductor $L_2$ is inserted within the lower branch of the feedback divider, enabling comparative testing of inductive loading directly inside the feedback path. This topology facilitates side-by-side investigation of input filtering, out-of-loop output LC filtering, and feedback-network inductance effects on transient response and small-signal stability.

The small-signal AC behaviour of the LDO with added inductors is summarized in Figure \ref{fig:ldo_ac_comparison}. Figure \ref{fig:bode_baseline} presents the baseline loop-gain characteristic of the unmodified LDO without an inductor; the response exhibits a smooth single-pole roll-off with adequate phase margin, confirming stable closed-loop operation. When the series inductor is inserted inside the feedback sensing path (Figure \ref{fig:bode_inside_loop}), the LC combination introduces a pair of resonant complex poles within the control loop. This manifests as pronounced gain peaking and a sharp deterioration in phase near the resonant frequency, severely degrading phase margin and producing the underdamped voltage ringing observed in transient simulations. In contrast, placing the inductor outside the feedback loop (Figure \ref{fig:bode_outside_loop}) confines the LC filter after the regulation sensing node. The control loop transfer function remains identical to the baseline case, and regulator stability is preserved, while the LC network still provides high-frequency noise filtering for the load. The output impedance comparison in Figure \ref{fig:zout_compare} further illustrates this distinction: a large resonant impedance peak emerges when the LC filter is embedded within the feedback loop, whereas the safe out-of-loop configuration avoids this problematic resonance. These results demonstrate that the physical placement of the inductor relative to the feedback divider is the dominant factor governing stability, rather than inductance magnitude alone.

\section{Conversation-based netlist generation}
\begin{figure}
\centering
\begin{subfigure}{0.23\textwidth}
\centering
\includegraphics[width=\textwidth, height=0.8\textwidth]{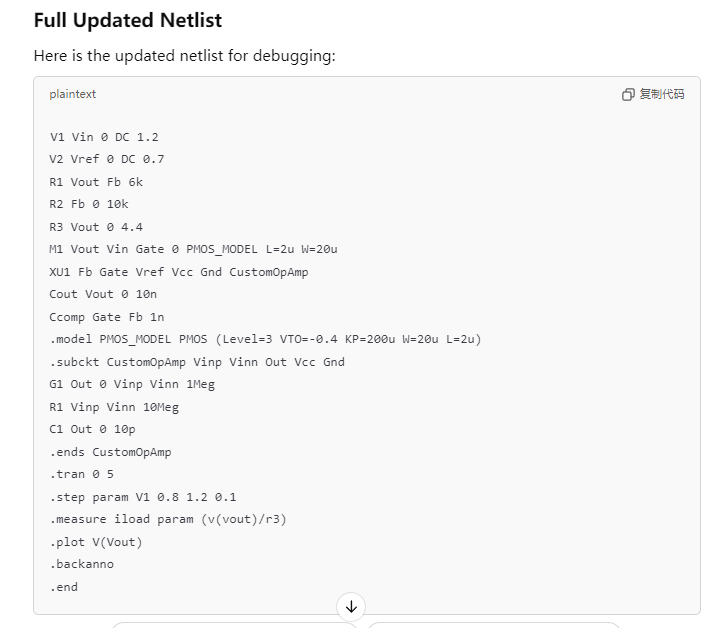}
\caption{AI generated Netlist}
\label{AI generated Netlist}
\end{subfigure}
\begin{subfigure}{0.23\textwidth}
\centering
\includegraphics[width=\textwidth, height=0.8\textwidth]{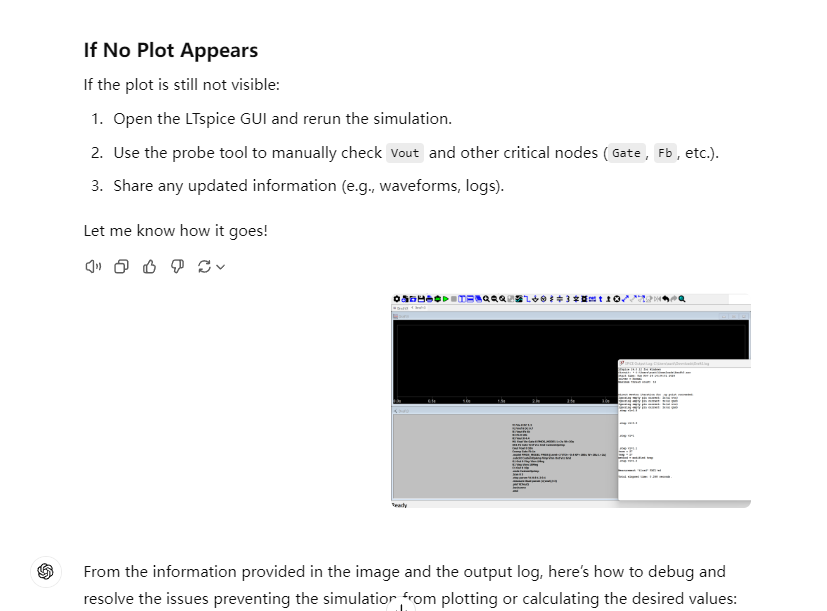}
\caption{The bug example}
\label{The bug example}
\end{subfigure}
\end{figure}

Netlists are a common way of getting circuits into a compact, textual form. Schematic diagrams can illustrate circuits in a more human-readable format, but for designs that more intricate it leads itself to a visual mess, thus Netlists, provide a more compact way to describe the components, the net connections between them, and how to simulate it (figure \ref{AI generated Netlist}). This makes it particularly good for LDOs vs other firms analog or mixed signal designs, which are often complex, as designers can focus on detailing process and become less entrenched into doing specific visual editing. An example is our design which is the use of netlists in LTspice to systematically build an LDO circuit with specifications for an input voltage of 0.8-1.2 V, an output voltage of 0.7-1.1 V, and a maximum load current of 250 mA. Everything from PMOS transistors, feedback resistors and compensation capacitors, to simulation settings — it was all defined explicitly through the Netlists, emphasizing the flexibility of this approach, even for advanced electronic designs.
Advantages of netlists in circuit design go beyond their textual nature. The biggest advantage is flexibility which allows the engineers to directly make targeted changes in the code without the need for redrawing charts as it would be complex to do so. Other than that, changing a resistor value or changing a node connection in a netlist is easy, and such alternatives are time saver if we are working on a repeated design. Finally, netlists promote portability and compatibility between various simulation tools, which is particularly beneficial in collaborative environments where different engineers may be using different tools. It is also a clear record of a circuit's design, helping with debugging because the relationships between components and their parameters are explicitly laid out. These attributes make netlists an essential resource in the field of electronics design—whether for industry professionals maximizing high-efficiency systems or students acquiring insights into circuit functionality.

In such context, the recent advent of AI in the field of circuit design has opened exciting avenues for automating netlist generation. AI can also read high level design specs and generate complete netlists from them, further decreasing manual input time. Generative AI netlists can ensure node interconnections and parameter assignments are accurate, either eliminating human error from the get-go or allowing designers to spend more time on higher-level architecture and optimization decisions. In addition, AI can greatly speed up iterative prototyping by quickly producing options certainly differing designs based on different parameters, something especially useful in performance-critical applications, such as LDO regulators. Not only does this increase overall productivity, but it also democratizes circuit design, enabling people with little technical expertise to engage with electronics.

Unfortunately, AI-generated netlists have significant shortcomings that need to be remedied. A big problem then is the contextual awareness deficiency of many AI systems, producing results that do not match specific design intents or atypical needs (figure \ref{The bug example}). For instance, in this project, when we tried to incorporate a custom op-amp subcircuit, the AI produced syntax errors and erroneous connections, indicating a failure to deal with something atypical or unconventional. Moreover, AI-generated netlists can be tricky to debug because the tools do not output enough information to find out what went wrong, forcing designers to retrace the error. This was illustrated early on when the simulation threw errors such as “measurement failed” and the expected output graphs, for example, Vout vs. Vin were not produced. These topics underscore that even though AI may godsend automate mundane tasks, machine human involvement is critical to shaping and verifying the output.
Looking back on this project, I think it really emphasized AI's current limitations of being able to generate netlists or debug them, but also potential for the future. However, AI played a key role in automating low-level and repetitive task and produced a netlist that was good enough in the first pass but needed to be edited. With the progression of AI, such techniques could very well be expanded to provide application-matched, context-familiar netlist generation and debugging capabilities that yield usable information. Additionally, dynamic optimization using the simulation results can improve the efficiency of AI tools. They are not quite reliable enough yet to handle all elements of circuit design, but their introduction into workflows represents an important leap forward. The future of AI in circuit design is one where the speed and efficiency of algorithms are coupled with the experience and knowledge of human designers to create an efficient and revolutionary design process.
\section{Conclusion}
This study demonstrates the application of GPT-4o in assisting with the design, simulation, and analysis of a low-dropout voltage regulator (LDO). GPT-4o has various suggestions in multiple stages of the design process, from brainstorming and proposing circuit topologies to guiding simulation setup and optimization. ChatGPT-4o will not directly feedback on the actual LDO circuit diagram, but it will feedback on the schematic of the LDO circuit diagram and the actual step based on the provided software. The biggest problem of ChatGPT-4o is the correctness check. It may miss the design specifications and will not be found by ChatGPT itself during the correctness check. The self-check is needed for the circuit design. However, ChatGPT can help the user to do accurate data distribution to satisfy the output requirement. In the design phase, GPT-4o was effective in suggesting a foundational LDO architecture, including key components such as a PMOS pass transistor, an error amplifier, and compensation elements. Notably, the operational amplifier design played a critical role in ensuring the stability and performance of the circuit. GPT-4o proposed a single-stage operational amplifier comprising a differential pair, current mirror, and buffer stage, with carefully chosen transistor dimensions for optimal performance.

Simulation results validated GPT-4o-assisted designs. In transient analysis, the LDO exhibited stable output voltage under varying input conditions, maintaining a dropout voltage within the design specifications. The addition of compensation capacitors significantly enhanced circuit stability, reducing the steady-state time. Frequency response analysis confirmed a stable phase margin and adequate gain bandwidth, ensuring robust performance across a wide frequency range. Furthermore, GPT-4o provided step-by-step guidance for conducting these simulations, making it a valuable tool for users unfamiliar with platforms like LTspice.

The operational amplifier, designed with thin-oxide NMOS and PMOS transistors, achieved high-speed operation and adequate gain. Its performance metrics, such as phase margin and DC gain, were validated through frequency response analysis, demonstrating that the design met stability requirements. The integration of a compensation capacitor further improved transient response, reducing the settling time and mitigating oscillations.

An investigation into integrated inductive magnetic elements reveals critical topology-dependent stability tradeoffs unique to linear LDOs, a layer of analysis also guided by iterative prompts to GPT-4o. Three distinct inductor placements are evaluated: an input supply filter inductor, an output LC filter positioned outside the feedback sensing tap, and an inductor embedded within the resistive feedback divider. Rooted in Maxwell’s electromagnetic induction laws and MOSFET small-signal amplifier theory, the analysis confirms inductors act as frequency-dependent impedance loads that introduce supplementary phase lag into the control loop only when placed inside the feedback path. The LC resonant pole pair formed by an inductor within the feedback divider triggers pronounced gain peaking, sharp phase degradation, and severely compressed phase margin, which manifests as sustained transient voltage ringing. By contrast, an identical inductor–capacitor LC network installed after the feedback sensing node exerts no influence over the core loop transfer function, fully preserving the original regulator stability while delivering high-frequency EMI filtering for the load. Component sizing analysis further identifies 47nH as the near-critically damped baseline inductance value for out-of-loop output filters, balancing noise attenuation and transient damping; inductances exceeding 96.8nH produce underdamped resonant behavior even in the stable external configuration. Input-side inductors solely isolate supply-borne interference without altering LDO regulation dynamics. This comparative study highlights that the relative position of magnetic inductive components with respect to the feedback divider dominates stability performance, rather than raw inductance magnitude, and underscores the necessity of small-signal AC verification when augmenting LDOs with LC passive filters—an important design caveat the LLM partially highlights but cannot fully quantify without user-specified simulation follow-up.

According to the experiment results we discussed above, we need to notice that, GPT-4o have some limits and disadvantages. For example, GPT-4o may miss a few components' data or configurations in a long conversation. With basic knowledge of circuit topologies and technologies and limited access to outer simulation tools like LTspice, GPT-4o failed to generate functionable Netlist for direct use. 

To a certain degree, GPT-4o can be a significantly useful tool for cricuit designers, especially amateurs, to brainstorm, construct, test, analyze, and optimzie the circuits. With appropraite and information-rich conversation, GPT-4o has a larger tendency to follow the correct path. However, while using the GPT-4o, users still need to pay attention to the possible missing context or midsleading responses. When realize there is any incorrect information or misleading response, users should give prompts that claim these mistakes and ask for double check and corrections. 

To sum up, ChatGPT-4o is not a perfect tool but still powerful. It is not long since the LLMs became popular, and LLMs from different development teams have been ungraded for many times. A large number of plug-in or extension tools for different LLMs have been created. We can optimically expect for more professional and specific field-oriented LLMs to be developed and improved in the near future.
\section{Declaration of competing interests}
The authors declare there is no competing interest.
\bibliographystyle{elsarticle-num}
\bibliography{ref}
\end{document}